# Origins of Optical Anisotropy in Nonspherical Plasmonic Particles: The Role of Surface Electron Scattering

Nikolai G. Khlebtsov[1*], Sergey V. Zarkov[1,2]

[1]Institute of Biochemistry and Physiology of Plants and Microorganisms, "Saratov Scientific Centre of the Russian Academy of Sciences," 13 Prospekt Entuziastov, Saratov 410049, Russia

[2]Institute of Precision Mechanics and Control, "Saratov Scientific Centre of the Russian Academy of Sciences," 24 Ulitsa Rabochaya, Saratov 410028, Russia

[*]To whom correspondence should be addressed. e-mail: (NGK) khlebtsov@ibppm.ru

ABSTRACT.

The optical anisotropy of non-spherical plasmonic particles, which arises from anisotropic geometric depolarization factors, is well established and enables tuning of dipole plasmon resonances across the UV–vis and NIR spectral ranges. However, a second type of optical anisotropy, caused by anisotropic surface electron scattering, remains considerably less explored. In contrast to the first mechanism, for which the dielectric function is isotropic, in the second the dielectric function of the nanoparticle itself becomes anisotropic because surface scattering contributes anisotropically to the damping constant. Although this mechanism emerges only within a quantum-mechanical description, a suitable representation can also be obtained within a generalized classical framework by using an anisotropic effective electron mean free path or, equivalently, an anisotropic surface scattering constant. Unlike depolarization anisotropy, surface scattering anisotropy shifts the plasmon resonance negligibly but can broaden it. This raises an important practical question: should this optical anisotropy mechanism be taken into account in simulations of absorption, scattering, metal-enhanced fluorescence, or surface-enhanced Raman scattering spectra? To date, this issue has not been discussed, and standard modeling practice ignores such anisotropy, relying instead on an isotropic approximation for the size-dependent dielectric function. Here, we address this problem using a quantum-mechanical framework (Uskov et al., Plasmonics, 2014) for gold and silver nanoparticles with three axial morphologies—rods, disks, and bicones (the latter as a model for experimental pentagonal bipyramids). In contrast to the scalar surface scattering constants commonly used for gold and silver nanoparticles, the electron scattering tensor depends strongly on the incident wavelength, particle shape, and orientation. Nevertheless, our main finding is that the resulting absorption and scattering spectra can be accurately predicted using a reasonable scalar approximation. For silver nanoparticles, we identify a universal plasmonic peak in the scattering tensor near 360 nm; its position is independent of particle shape and coincides with the resonance of the silver material function $1/|\varepsilon+1|$. We elucidate the plasmonic mechanisms underlying this phenomenon

## 1. INTRODUCTION

The size-induced broadening of plasmon resonances in metallic nanoparticles has been recognized since the pioneering absorption measurements in 1958 reported by Doyle [1] on sodium nanoparticles embedded in NaCl crystals, the experiments of Hampe [2] on silver nanoparticles in a silicate glass matrix, and Fragstein and Römer [3] on gold sols. The resonance widths observed by Doyle and Hampe were found to be substantially larger than the theoretical predictions based on the classical Drude model employing the bulk dielectric function. This discrepancy was ascribed by Doyle and Hampe to the confinement of the electron free path imposed by the finite particle dimensions. Specifically, the imaginary part of the dielectric response of plasmonic nanoparticles was modelled by invoking a plasma resonance of the conduction electrons within individual metal nanospheres, with the damping of the collective oscillations being attributed solely to Ohmic losses, the magnitude of which was governed by the free-path limitation due to the sphere radius. The above physical picture was consistent with Euler's [4] earlier discussion of certain anomalies in the optical properties of thin gold films, explained in terms of the limited mean free path of electrons in metal spheroids. Three years thereafter (in 1961), Römer and Fragstein [5] reported a strong size-dependent variation of the real part of the refractive index of gold nanoparticles in the size range of 3–4 nm. Referring to the studies of Doyle [1] and Hampe [2], they likewise interpreted this anomalous behavior in terms of the restricted electron free path. Within the framework of this conventional approach, Kreibig and Fragstein [6] proposed the following size-corrected expression for the damping constant $\gamma_a$ in a spherical particle of radius $a$

$$\gamma_a = \gamma_b + \gamma_s = \gamma_b + A_s \frac{v_F}{a}, \tag{1}$$

where $\gamma_b$ is the damping constant of bulk metal, $\gamma_s$ describes the surface scattering of electrons, and $v_F$ is the Fermi velocity, and the phenomenological constant $A_s$ accounts for the effects of surface potential, electron spillover, dielectric environment, and typically is close to 1. This model is sometimes referred to as the classical scattering (CS) model [7], which is discussed in detail by Kreibig and Vollmer [8]. In general, an additional chemical interface damping (CID) $\gamma_{CID}$ [9, 10] can be added to (1) or included into the scattering constant $A_s$. The modified damping constant (1) should be used in the Drude expression for the dielectric function of free electrons. To include the interband contribution to the dielectric function of bulk metals $\varepsilon_b(\omega)$, the following procedure is typically used [11]:

$$\varepsilon(\omega, a) = \varepsilon_b(\omega) + \Delta\varepsilon(\omega, a), \tag{2}$$

where the correction $\Delta\varepsilon(\omega, a)$ accounts for the contribution of size-dependent electron scattering to the Drude part of the dielectric function

$$\Delta\varepsilon(\omega, a) = \frac{\omega_p^2}{\omega(\omega + i\gamma_b)} - \frac{\omega_p^2}{\omega(\omega + i\gamma_a)}, \tag{3}$$

$\omega_p$ is the plasma frequency. Starting from these pioneering works, the classical concept of the size-corrected electron mean free path in nanoparticles has been successfully used in numerous studies [11-14] (see also related references in the reviews [15, 16]).

As pointed out by Kawabata and Kubo [17], the above classical description of the appearance of size-dependent damping due to electron scattering at the surface is not entirely correct. From a quantum-mechanical (QM) point of view, a reduction in particle size below 3–5 nm leads to the emergence of discrete surface levels [18], and the appearance of an additional contribution to the imaginary part of the particle's dielectric function is due to dipole transitions between discrete surface levels. Despite a completely different physical mechanism, the QM calculations performed by Kawabata and Kubo, as well as by other authors [19, 20], have in fact led to the same formula (1) with a scattering coefficient of the order of unity.

The obvious drawback of the classical approach (1) is that it does not account for the anisotropy of particle shape and the orientation of the field relative to the particle. For spherical particles, this does not pose any problems, since the particle radius is the only characteristic size, and the damping constant (1) is a scalar. Coronado and Schatz [21] generalized (1) to the case of non-spherical particles, for which the radius of the spherical particle is replaced by an effective mean free path

$$l_{eff} = 4V / S , \tag{4}$$

where $V$ and $S$ are the particle volume and surface area. In fact, this generalization also employs a single average scalar parameter for electron scattering. However, it is intuitively clear that for particles such as nanorods, there are two characteristic sizes and, accordingly, two effective electron mean free paths. In other words, the damping constant (1) and the corresponding dielectric function should be anisotropic and take the form of tensors [22] (axial in the case of nanorods or disks). But the classical approach does not provide a recipe for how such tensors can be obtained. Arguably, an exception is the work [23, 24], in which the classical kinetic Boltzmann equation was employed to calculate the anisotropic conductivity of nanospheroids with dimensions much smaller than the electron mean free path. This tensor conductivity is related in a known manner [24] to the tensor of the optical dielectric permittivity $\varepsilon_{ij} = \delta_{ij} + i4\pi\sigma_{ij} / \omega$.

In quantum-mechanical (QM) approaches [8, 20, 22, 25-30], size and shape effects can, in principle, be incorporated, although analytical estimates have been obtained only for particles of simple geometries, such as spheres, rectangular prisms, cylinders, and spheroids, for which the spectrum of electronic states can be calculated. Uskov et al. [31] derived an analytical QM generalization of the scalar relation (1) to the case of spheroids. In their theory, the scalar scattering constant is replaced by two constants corresponding to the field orientation parallel and perpendicular to the particle's rotation axis (these results were also reproduced in [7]). It was shown that for oblate spheroids, the scattering constant increases with increasing aspect ratio (diameter-to-thickness ratio).

Thus, both the intuitive classical approach and the QM analysis clearly indicate that for non-spherical plasmonic particles, the size correction should be anisotropic in nature, leading to an anisotropic damping constant and, consequently, an anisotropic dielectric function already at the Drude level. In its clearest form, the anisotropic character of the size-dependent dielectric function and related plasmon resonance broadening was formulated by Kraus and Schatz [22], who calculated the size-dependent contribution to the width for several particle shapes and field directions (Table 1 in Ref. [22]). Nevertheless, current practice in plasmonics [32-35] tends to disregard this issue, and calculations of the optical properties of non-spherical plasmonic particles are performed using either the bulk dielectric functions or an isotropic effective size parameter according to the Coronado–Schatz prescription [21, 36], or the radius of an equivalent-volume sphere [37]. This raises an important question: to what extent is such a practice of neglecting the anisotropy of the dielectric function, arising from the anisotropy of the damping parameter, justified? How significant could the differences be, for instance, in the absorption and scattering spectra of non-spherical plasmonic particles when computed with due account of anisotropy as opposed to the currently prevailing practice in plasmonics of employing an isotropic size correction to tabulated bulk data? To the best of our knowledge, such questions have not even been posed so far, let alone answered. Among the few exceptions, one can only mention the work of Govorov and co-workers [38, 39] on the modeling of hot-electron properties in nanorods and nanostars and the functional density simulations of nanometric gold thin films [40, 41].

In this work, we investigate the effect of the anisotropy of the damping parameter on the extinction, absorption, and scattering spectra of non-spherical gold and silver nanoparticles, including nanorods, disks, and bicones as a model of pentagonal bipyramids (Supporting Information (SI) file, Figure S1). We identify the regimes in which the commonly adopted practice of neglecting anisotropy and employing isotropic size-corrected optical constants of the metal is justifiable, as well as those cases where more rigorous models may be required. Although our analysis is restricted to scattering and absorption spectra, it is fully applicable to other problems, including the modeling of SERS spectra and metal-enhanced fluorescence (MEF).

In concluding this section, we emphasize the distinction between two origins of the optical anisotropy in plasmonic particles:

(1) the anisotropy of the polarizability inherent to any non-spherical particle with an isotropic dielectric function, corresponding either to an isotropic bulk material or to an isotropic size-corrected response. This contribution primarily affects the *position* of resonances (e.g., transverse and longitudinal dipolar modes), while having little effect on the linewidth broadening for particle sizes larger than 10 nm.

(2) the anisotropy of the dielectric function itself, arising from the geometric anisotropy of the electron scattering mechanism [22, 24]. This mechanism has virtually no effect on the resonance position, but induces *additional anisotropic broadening* of the spectral lines, as compared with the broadening due to other (intrinsic) mechanisms (e.g., phonon and impurity scattering).

In the former case, the optical anisotropy arises (within the dipole approximation) from the anisotropy of the depolarization factors of non-spherical particles composed of an isotropic material. This gives rise to anisotropy in the polarizability and optical responses, including UV-vis spectra, SERS, MEF, orientational dichroism and birefringence [42], depolarization of scattered light [43-45], etc. In the present work, electron scattering at the surface of a non-spherical particle leads to an intrinsic anisotropy of the particle material itself due to the surface scattering tensor $A_s^{ij}$, which should replace the scalar scattering coefficient $A_s$. Consequently, the anisotropy of the dipolar polarizability stems not only from the anisotropy of the geometrical depolarization factors $L_i$ [46], but also from the anisotropy of the material itself. This becomes evident if one writes down the standard expression [46] for the polarizability of a small ellipsoid along its principal symmetry axes $i = x, y, z$:

$$\alpha_i = \frac{V}{4\pi} \frac{\varepsilon_i - \varepsilon_m}{\varepsilon_m + L_i(\varepsilon_i - \varepsilon_m)}, \tag{5}$$

where $\varepsilon_m$ is the dielectric function of the environment. The principal values of the dielectric function tensor $\varepsilon_i$ are determined by the principal values of the damping tensor $\gamma_i$.

## II. THEORETICAL METHODS

### A. Two approaches to the parametrization of size effects in nanoparticles

Referring to equation (1), two approaches are possible for describing the dependence of the damping parameter on nanoparticle size. In the first approach, a certain characteristic particle size $l_{nano}$ is fixed [31], while all other details of the electron scattering mechanism, including the dependence on shape, field direction, etc., are incorporated into the scattering coefficient $A_s$. In the second approach, the scattering coefficient $A_s$ in equation (1) is fixed and is usually set equal to unity, whereas all remaining details are accounted for by an effective mean free path [20]

$$l_s = \gamma_s / v_F . \tag{6}$$

In this context, Kraus and Schatz [31] employed as the characteristic particle size an average length along the applied field

$$l_{av} = V / S_{\perp}, \tag{7}$$

where $S_{\perp}$ is the projected area of the particle perpendicular to the direction of the applied field. An important result of the quantum-mechanical calculations is that for several simple particle shapes (sphere, cylinder, and prism) the ratio $l_s / l_{av} \simeq \mathrm{const} \approx 0.65$. It should be emphasized that for nonspherical particles, both the scattering coefficient and the effective length are, in general, tensors. In the present work, we adopt the first approach, using as the characteristic size the radius of the volume-equivalent sphere

$$a_0 \equiv R_{ev} = \sqrt[3]{3V/4\pi} \ . \tag{8}$$

The same characteristic size $R_{eff} = \sqrt[3]{3V/4\pi} \equiv a_0$ was used in Ref. [38] (see their Eq. (24)). For comparison, we shall also employ the average effective size defined by equation (4). In terms of these sizes, the contribution of surface scattering to the damping constant is given by the Kreibig formula

$$\gamma = \gamma_b + \gamma_s = \gamma_b + A_s \frac{v_F}{a_0} = \gamma_b + A_{eff} \frac{v_F}{l_{eff}} , \tag{9}$$

where $\gamma_s = A_s \frac{v_F}{a_0} = A_{eff} \frac{v_F}{l_{eff}}$, so that the scattering constants differ only by the factor $A_{eff} / A_s = 4V / a_0 S = 16 a_0^2 / 3S$ . For a sphere, $l_{eff} = 4a_0 / 3$, hence $A_{eff} / A_s = 4/3$ .

**B. Quantum mechanical surface-scattering parameter**

The quantum mechanical surface-scattering damping parameter [31] can be written in the following form:

$$\gamma_s^{\mathbf{e}} = A_s^{\mathbf{e}} \frac{v_F}{a_0} , \tag{10}$$

$$A_s^{\mathbf{e}} = F_m(\omega) H_g^{\mathbf{e}}(\omega) , \tag{11}$$

$$H_g^{\mathbf{e}}(\omega) = \frac{a_0 \int |\mathbf{n}\mathbf{E}_i|^2 \, \mathrm{d}S}{\int |\mathbf{E}_i|^2 \, \mathrm{dV}} , \tag{12}$$

where the superscript corresponds to a specific polarization $\mathbf{e}_0$ of the incident light, the dimensionless *geometric* factor $H_g^{\mathbf{e}}$ is determined by the particle shape and the orientation of the incident field with respect to the particle, $\mathbf{n}$ is the unit normal vector at a given point on the inner surface of the particle, and $\mathbf{E}_i$ is the field inside the particle. The material parameter $F_m$ depends on the particle material and on the frequency (wavelength) of the light.

$$F_{mat}(\omega) = C_{mat} G(x = \hbar\,\omega / E_F), \tag{13}$$

$$C_{mat} = \frac{4\alpha}{\pi}\frac{c}{v_F}\left(\frac{E_F}{\hbar\omega_p}\right)^2 . \tag{14}$$

where $\alpha = \frac{e^2}{\hbar c} \simeq \frac{1}{137}$ is the fine structure parameter, $E_F$ is the Fermi energy. The function *G(x)* is defined by integrals (10) of Ref. [47] (SI File, Section S2) that can be taken analytically:

$$G(x) = \frac{2+3x}{6x}r^3 - \frac{2+x}{4}\left[r + xp\right] - \frac{(1-x)}{2}\left[\frac{y(2-x)}{x} - xq\right] + \frac{2y^3}{3x} - \frac{y(2-x)}{4} + \frac{qx^2}{4}, \tag{15}$$

where $r = \sqrt{1+x}$, $y = \sqrt{1-x}$, $p = \ln\left((1+r)/\sqrt{x}\right)$, and $q = \ln\left((1+y)/\sqrt{x}\right)$. Note that another analytical expression for $G(x)$ was provided in Ref. [31]

$$G(x) \equiv G_1(x = \hbar\omega / E_F) = \frac{1}{3x}\left[1 - \left(1-x\right)^3\right]. \tag{16}$$

Both functions have equal low-energy limits $G(0) = G_1(0) = 1$, but different behavior in the range $0 \le x \le 1$. Actually, the function (16) is valid only in the low-energy limit $x \ll 1$, whereas the function (15) works for all photon energies below the Fermi energy (Figure S2, SI file).

Upon employing the effective length and the corresponding scattering constant, one finds that

$$A_{eff}^{\mathbf{e}} = F_m(\omega) H_{eff}^{\mathbf{e}}(\omega), \tag{17}$$

$$H_{eff}^{\mathbf{e}}(\omega) = \frac{\frac{4}{S}\int \left|\mathbf{n}\mathbf{E}_i\right|^2 \mathrm{d}S}{\frac{1}{V}\int \left|\mathbf{E}_i\right|^2 \mathrm{dV}}, \tag{18}$$

It is readily seen that for spheres $H_g^{\mathbf{e}} = 1$, whereas the effective constant $H_{eff}^{\mathbf{e}} = 4/3$. However, the effective scattering length for spheres is $l_{eff} = 4a/3$, so that the physical quantity, namely the surface damping constant $\gamma_s = F_m \frac{v_F}{a}$, remains invariant, as is required. Kraus and Schatz [22] found for spheres a slightly smaller value $l_{eff} \simeq 1.16$, whereas Ruppin and Yatom [25] reported a somewhat higher value of $l_{eff} \simeq 1.64$.

By comparing our Eqs. (10)-(12) with similar relations given in Refs. [7, 38] we note the following difference: a constant multiplier 3/4 stands in [7, 38] instead of our material factor $F_m$ (Eq. (13)). Specifically, the surface scattering damping constant is expressed in [38] as $\gamma_s = b(\omega)\left(v_F / R_{eff}\right)$ with $R_{eff} = R_{ev} \equiv a_0$ and $b(\omega) = (3/4)H_g^{e}(\omega)$. In other words, the material factor $F_m(\omega)$ in our Eq. (14) is replaced in [38] by a constant multiplier 3/4. Further, for both gold and silver particles, Eq. (14) gives $C_{mat} \simeq 0.75 = 3/4$, and in the low-energy limit ($\hbar\omega / E_F \ll 1$), we obtain $G(x) \simeq 1$. Thus, we arrive at $A_s = F_m \simeq 3/4$, which agrees with [7, 27, 38]. Note, however, that such agreement is observed only in the low-energy limit, while in the general case the material factor depends on the wavelength of light and on the material parameters (Figure S2, SI file).

For small metal spheroids, the inner field is a homogeneous one, so the geometrical factor (15) is calculated analytically [31] (Section S3, Figure S3, SI file). In agreement with previous data [31], the geometrical factors strongly depend on the aspect ratio of spheroids. Most significant anisotropic deviations from the spherical limit $H_s^{z,x} = 1$ are observed for oblate particles.

If the external field is oriented at an arbitrary angle $\theta = \pi/2 - \vartheta$ ($\vartheta$ is the angle between the wave vector and the symmetry axis $c$ of a small spheroid, Figure S6), then, by virtue of the superposition principle, the internal field remains homogeneous [48], just as in the case where the external field is aligned along either of the symmetry axes. However, it will not be parallel to the external field owing to the anisotropy of the depolarization factors, $\mathbf{E}_i = \hat{p}\mathbf{E}_0$, where the axial tensor $\hat{p}$ has two principal components along $p_z$ and perpendicular $p_x$ to the symmetry axis

$$p_{x,z} = \frac{1}{1 + L_{x,z}\left(\varepsilon_r - 1\right)}, \tag{19}$$

where $\varepsilon_r = \varepsilon / \varepsilon_m$ and $L_{x,z}$ are the well-known geometrical depolarization factors of spheroids [46, 48]. By using the general expression (12), we derived the following analytical solution for the geometrical factor $H_g(\vartheta)$ (Section S3 in the SI file)

$$H_g^e(\vartheta) = \frac{H_g^x p_x^2 \cos^2\vartheta + H_g^z p_z^2 \sin^2\vartheta}{p_x^2 \cos^2\vartheta + p_z^2 \sin^2\vartheta}. \tag{20}$$

Thus, the $H_g(\vartheta)$ is expressed as a linear combination of $H_g^x$ and $H_g^z$ with simple analytical angle-dependent coefficients as defined by Eq. (20).

Since the field inside the particle depends on the dielectric function, the right-hand side of equation (10) is a function of the damping parameter. The functional equation (10) can be solved by iterations [38, 39], the convergence of which depends on the particle parameters and the nature of the metal (for silver, the convergence is worse than for gold). Figure S4 (SI file, Section S4) gives an example of the convergence of iterations. To reduce the number of successive iterations, the analytical geometrical factors for spheroids of equal diameter and aspect ratio were used as the initial guess (whereas in Ref. [38], they were simply set to zero).

**C. Numerical simulations**

For numerical simulations, we used the commercial package COMSOL Multiphysics 5.1 (further COMSOL or COM (in Figures) for sake of brevity) in both its standard 3D and a novel 2.5D [35, 49] options. The latter is applied to particles with axial symmetry by expanding all fields in cylindrical harmonics and calculating each contribution independently. The main advantage of the 2.5D approach is its high computation speed, which makes random orientation averaging feasible within an acceptable CPU time. For isotropic particles with isotropic surface scattering correction, we employed T-matrix codes (TMM) [50], which provide the fastest calculations of orientation-averaged electromagnetic cross sections, used as a benchmark for comparison with the COMSOL data.

In practice, most extinction spectra measurements are performed on colloidal solutions in which the plasmonic particles have random orientations. The corresponding theoretical quantity is the orientation-averaged extinction and scattering cross sections that should be calculated by integration over three Euler angles $(\alpha, \beta, \gamma)$ with a constant probability density function $p(\alpha, \beta, \gamma) = 1/8\pi^2$ (see Section S5 in the SI file). Numerical orientation averaging is often performed using three particle orientations along the coordinate frame axes and two polarizations of light propagating along one axis. Alternatively, one can use various configurations of the incident light, then sum all cross sections and divide by the number of configurations (see, e.g., Eq. (8) in Ref. [34], Eq. (6) in Ref. [51], and Section S5 in the SI file). It should be emphasized that such averaging is acceptable for particles with a random structure, like nanostars, but it can lose some important resonances for highly symmetrical particles, such as nanorods. Figure S7 illustrates inaccurate orientation averaging of the normalized extinction cross section using several particular configurations of the incident light (designated as $\mathrm{COM}_{\mathrm{av}}$) in comparison with the exact orientation averaging solutions based on the general T-matrix approach for arbitrarily shaped particles [52, 53]

$$\langle C_{ext}\rangle = -\frac{2\pi}{k^2}\mathrm{Re}\left[\mathrm{Tr}(\hat{T})\right],\ \langle C_{sca}\rangle = \frac{2\pi}{k^2}\left[\mathrm{Tr}(\hat{T}^{+}\hat{T})\right] = \frac{2\pi}{k^2}\left\|\hat{T}\right\|^2. \qquad (21)$$

For simplicity, we considered randomly oriented silver nanorods with bulk optical constants. Similar results were obtained using size-corrected optical constants. The physical origin of the loss of the quadrupole peak at 568 nm is that the most effective excitation of this quadrupole occurs when the particle axis is oriented at 54 degrees with respect to the incident light [54]. Obviously, none of the fixed orientations along the coordinate frame associated with the incident light can excite the quadrupole peak.

Although the orientation averaging requires numerical integration over three Euler angles, for axially symmetric particles it is sufficient to consider only TE ($\mathbf{E}\perp(\mathbf{k},\mathbf{c})$) and TM ( $\mathbf{H}\perp(\mathbf{k},\mathbf{c}),\mathbf{E}\in(\mathbf{k},\mathbf{c})$) configurations of the incident light, where $\mathbf{c}$ is the particle symmetry axis (Figure S5). The orientation-averaged extinction cross section is then given by

$$\langle C_{ext}\rangle = \int_0^{\pi/2}\sin\vartheta d\vartheta \frac{1}{2}\left[C_{ext}^{TE}\left(\vartheta, A_s^{TE}(\vartheta)\right) + C_{ext}^{TM}\left(\vartheta, A_s^{TM}(\vartheta)\right)\right], \qquad (22)$$

where $\vartheta$ is the angle between the vectors $\mathbf{k}$ and $\mathbf{c}$. Note that $A_s^{TE,TM}(\vartheta)$ should be calculated by solving Eqs. (17), (18) for each orientation. For small nanorods, $C_{ext}^{TE}(\vartheta)$ weakly depends on the angle $\vartheta$, and $C_{ext}^{TE}(0)\simeq C_{ext}^{TE}(90)$ for propagation along and perpendicular to the symmetry axis, respectively (Section S5, Figure S6 in the SI file). Thus, Eq. (22) can be replaced by

$$\langle C_{ext}\rangle = \frac{C_{ext}^{TE}\left(0, A_s^{TE}(0)\right) + C_{ext}^{TE}\left(90, A_s^{TE}(90)\right)}{4} + \frac{1}{2}\int_0^{\pi/2}\sin\vartheta d\vartheta\left[C_{ext}^{TM}\left(\vartheta, A_s^{TM}(\vartheta)\right)\right] \qquad (23)$$

A similar relation also holds for the scattering and absorption cross sections.

The bulk dielectric function of gold was calculated by a spline interpolation of tabulated data [55]. For bulk silver, we used the spline described in Section S9 of the SI file (**Figure S12).**

## II. RESULTS AND DISCUSSION

### A. Gold and silver nanorods

For metal nanospheroids, the internal field is homogeneous, and the geometric factors, as defined by Eqs. (12) and (15), do not depend on the particle volume or the frequency of light. Thus, the aspect ratio is the only shape parameter that determines the surface damping constant. Similarly, for plasmonic nanorods, the scattering constant and the corresponding surface damping constant (i.e., the surface scattering rate [7]) also depend only weakly on the particle volume [38]. However, in contrast to spheroids, the internal field in nanorods is not homogeneous, and the geometric factors depend not only on the particle aspect ratio but also on the light frequency. As noted above, an additional spectral dependence also arises from the material factor in Eq. (11). Therefore, all calculations were performed with a fixed nanorod diameter of 10 nm (a typical value for most biomedical applications [56]), while the aspect ratio was varied from 2 to 6. Such geometrical parameters coincide with those used in Ref. [39].

Figure 1 shows the spectral dependence for two principal values of the surface scattering tensor $A_s^z$ and $A_s^x = A_s^y$ corresponding to excitation of rods along the symmetry axis $z$ and in perpendicular directions $(x, y)$. Figure S8 demonstrates differences in spectral dependence of the principal components of the scattering tensor for nanorods and spheroids with similar geometrical parameters. Here and in what follows, the symmetry axis $\mathbf{c}$ of an axial particle is supposed to be directed along the $z$ axis of the particle coordinate frame. The calculation results lead to the following conclusions:

1. The scattering constants for the longitudinal and transverse field orientations exhibit strong anisotropy, differing by nearly an order of magnitude in the case of elongated nanorods (cf. the spectra in panels A and B).
2. In contrast to prolate spheroids with a homogeneous internal field (Figure S8), the inhomogeneous field distribution within nanorods gives rise to a pronounced wavelength dependence of the scattering coefficients in the short-wavelength region. The spectral peaks are localized near the transverse plasmon resonances (~370 nm for silver and ~500 nm for gold). These results agree with previous simulations for gold and silver nanorods [38, 39].
3. For silver nanorods, the spectral peak position is independent of the aspect ratio; however, no such aspect-ratio independence is observed for gold nanorods.
4. Contrary to the findings reported in [33, 34], the scattering coefficients for the transverse field orientation display a weak increasing, rather than decreasing, dependence on wavelength ( $\lambda \geq 400$ nm). This discrepancy is attributable to the presence of a spectrally dependent quantum-mechanical material factor in equation (11), as opposed to the constant prefactor 3/4.
5. Minor differences are observed in the scattering coefficient spectra upon transverse-field excitation, depending on whether the wave vector is oriented perpendicular or parallel to the symmetry axis (configurations denoted by circles and asterisks, respectively). These differences arise from slight deviations from the electrostatic limit, in which the aforementioned configurations are equivalent. Such deviations become more pronounced with increasing particle size.
6. For nanorods with an aspect ratio exceeding 4 and at wavelengths longer than 600 nm, the differences between the coefficient spectra calculated with the material factor (equation (11)) and those obtained with the prefactor ¾ are negligible (cf. the solid and dashed curves in the insets to panels A and C).

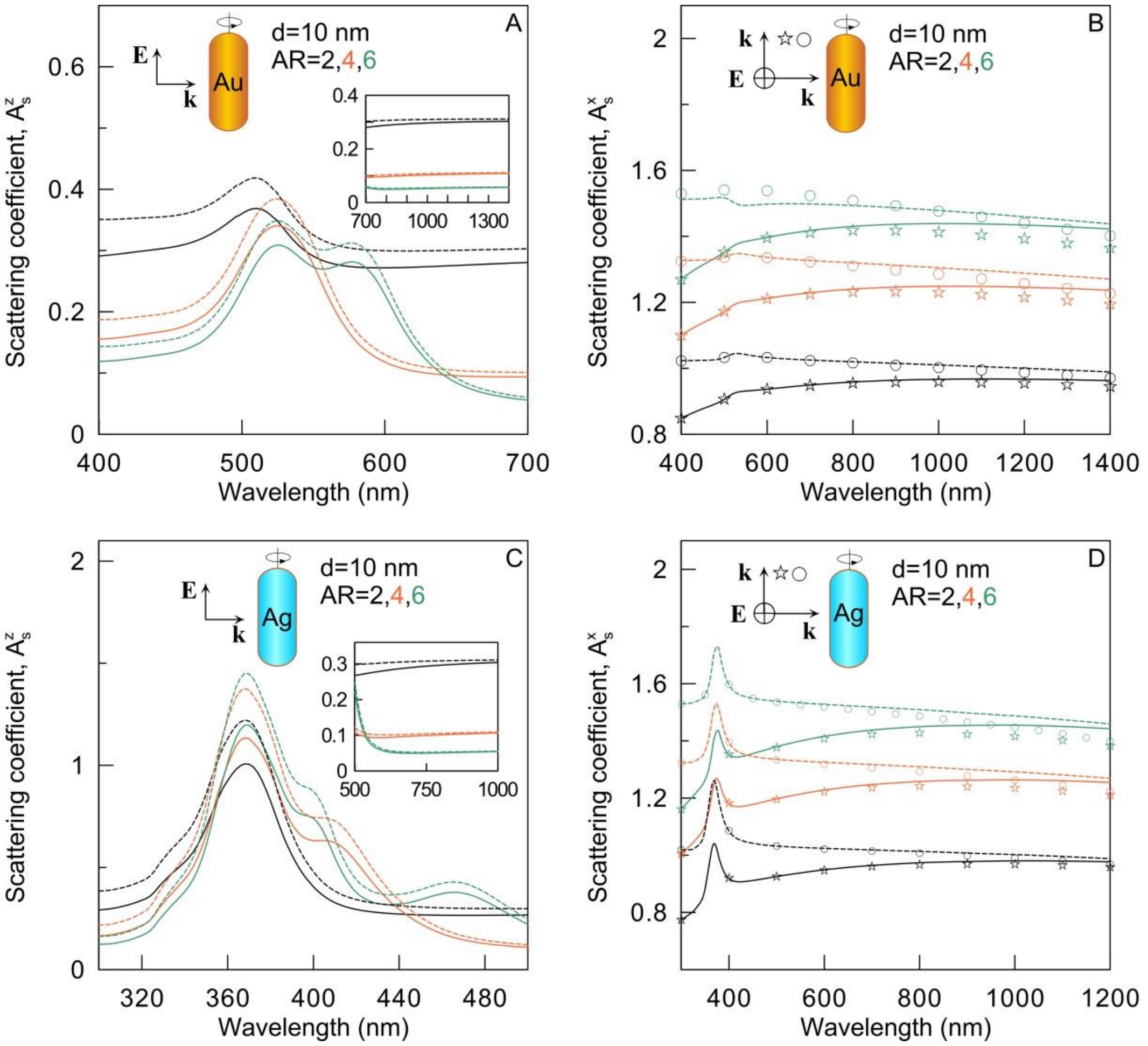

**Figure 1**. Spectra of the scattering coefficients $A_s^{z,x}$ (the principal values of the $A_s^{ij}$ tensor) for gold (A, B) and silver (C, D) nanorods with diameters 10 nm and aspect ratios of 2 (black), 4 (orange), and 6 (green). For comparison, the dashed lines show the scattering coefficients $(3/4)H_g^{z,x}$ [7, 38]. The insets in panels A and C illustrate small differences between the scattering coefficients $A_s^z$ and $(3/4)H_g^{z,x}$ for wavelengths above 800 (A) and 600 nm (C). In panels C and D, circles and asterisks show calculations for the TE0 ($\mathbf{E}\perp\mathbf{z};\mathbf{k}\,||\,\mathbf{z}$)configuration, while the solid and dashed lines show calculations for the TE90 ($\mathbf{E}\perp\mathbf{z};\mathbf{k}\perp\mathbf{z}$) configuration.

A precise calculation of orientation-averaged cross sections requires knowledge of how the scattering constant depends on the angle of the nanorod's orientation relative to the incident wave. Equation (20) provides an approximate solution to this problem for spheroids, offering insight into the key trends in the angular dependence of the geometric factor that governs the angular dependence of the scattering constant. Since the cross sections for the TE configuration are virtually independent of the rod's orientation, only the TM configuration needs to be considered.

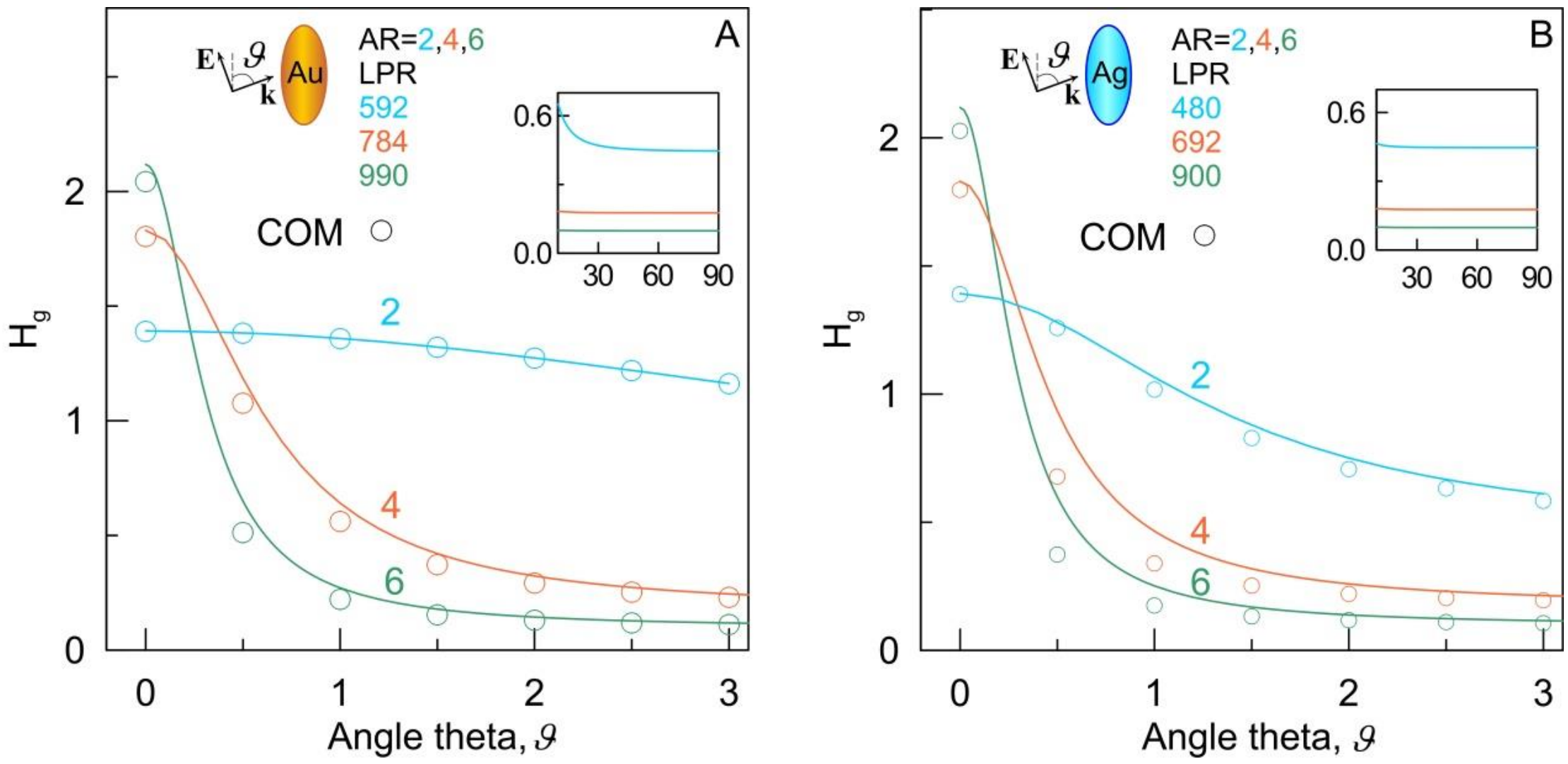

**Figure 2.** Dependence of the geometrical factor $H_g^z$ on the orientation of gold (A) and silver (B) spheroids with respect to the wave vector for TM configurations. The solid lines show analytical calculations by Eq. (20) for aspect ratios of 2 (blue), 4 (orange), and 6 (green). The corresponding LPR wavelengths are indicated in the panels. The circles show COMSOL calculations for comparison. The insets show a weak angular dependence for $\vartheta$=10–90°, where the analytical and numerical data coincide. All calculations are performed for a particle diameter of 10 nm using the bulk dielectric functions.

Figure 2 presents the angular dependences of the geometrical factor $H_g^z$ for gold (A) and silver (B) spheroids, calculated analytically using Eq. (20) and numerically using COMSOL and the bulk optical constants. First, we note the good accuracy of the analytical approximation compared with the exact numerical calculation. Second, for aspect ratios of 4 and 6, a sharp decrease in the geometrical factor by one order of magnitude is observed within a narrow range of orientation angles between zero and three degrees. For all other angles, the geometrical factors are virtually constant and close to the value at 90 degrees (the TM90 configuration). Similar results were obtained for the more complex case of gold and silver nanorods, taking into account the iterative solution for the anisotropic scattering constant. Figure 2 shows that, with such a strong dependence of the scattering constant on the orientation angle, calculating average cross sections by summing over several incident-wave configurations cannot reveal the influence of the angular dependence of the scattering constant on the average cross-sections. In particular, simple summation of the configurations for a zero incidence angle does not take into account that these maximum scattering constants enter the summed cross sections with a weight $\sin\vartheta$ equal to zero.

Figure 3 presents the extinction and scattering spectra of gold and silver nanorods calculated for excitation along the axis of symmetry (panels A, B and D, E), as well as the spectra of particles with random orientation (panels C and F). The green lines show calculations taking into account the anisotropy of the electron scattering constant, which depends on the wavelength and the orientation of the nanorods. A model with a constant scattering coefficient $A_s = 0.75$, to which the quantum-mechanical calculation for spherical particles reduces, was chosen as a reference. This value (along with 1) is also frequently used for modeling the spectra of gold and silver nanoparticles with allowance for the size limitation of the electron mean free path. Additionally, Figure S9 presents, for comparison, a calculation with a constant scattering parameter $A_s = 0.25$ obtained by fitting experimental data for single spheres [57] and nanorods [58]. In ref. [58], $l_{eff} = d\left[1-1/(3AR)\right]$ according to Eq. (4) was used as the characteristic scattering length, whereas for the spherical particles studied in the experiments of ref. [56], $l_{eff}^{*} = 4R/3 = 2d/3$ was used; therefore, the equivalent $A_s^{*} = 0.25 \times l_{eff} / l_{eff}^{*} \simeq 0.33$. This value was also obtained by fitting data for polydisperse nanorods [59] using T-matrix orientational averaging over a large ensemble of TEM images (on the order of 1000 particles).

The spectra calculated with allowance for the anisotropy of the scattering constant prove, somewhat unexpectedly, to be fairly high in quality. In terms of the magnitude of the maximum and the half-width, they are closer to the spectra obtained with bulk constants than to the spectra obtained with the frequently used constants of 0.75 or 1 [16]. In this connection, we note the data of ref. [60], in which agreement between the scattering spectra for single nanorods with an aspect ratio was obtained for an unusually small scattering constant $A_s = 0.13$. To account for this, the authors employed the value $A_s = 0.3$ obtained by Persson [61] from a quantum-mechanical calculation for spherical particles. In order to apply this estimate to rods, the authors of ref. [60] used the assumption of Kreibig and Vollmer [8] that for nanorods the major semiaxis, whose length determines the position of the longitudinal resonance peak, should be chosen as the effective length. The T-matrix modeling of the spectra in ref. [59] was performed for an approximate model of spheroids [62] with the radius of the equivalent-volume sphere $R_{ev}$ serving as the characteristic length for the size correction of the bulk optical constants of gold. Since this size is smaller than the major semiaxis of the spheroid $L/2$, the optical constant $A_s'$ used for the calculation must be reduced [58] by a factor $q = A_s'(R_{ev}) / A_s(L/2) = AR^{-2/3}$. In our simulations, we used a model of cylinders with hemispherical ends, which is more adequate to the shape of experimental gold nanorods [37]. For this model we obtain $L = d \times AR$, $R_{ev} = (d/2)\sqrt[3]{(3AR-1)/2}$, and

$$A_s^*(R_{ev}) = qA_s(L/2), \; q = \frac{R_{ev}}{L/2} = \frac{\sqrt[3]{(3AR-1)/2}}{AR}. \qquad (24)$$

For our values $AR = 2$, 4 and 6, the parameter $q =$0.679, 0.467, and 0.340. Following [60, 61], we assume $A_s = 0.3$, which gives $A_s^* \simeq$0.2, 0.13, and 0.1. For gold nanospheres, Ruppin and Yatom [25] obtained $A_s = 6g(\hbar\omega / E_F)/\pi^2 \simeq 0.6 - 0.4$ (400-900 nm). For silver spheres in a glass matrix, the measurements by Gentzel et al. [20] give $A_s \simeq 0.9$.

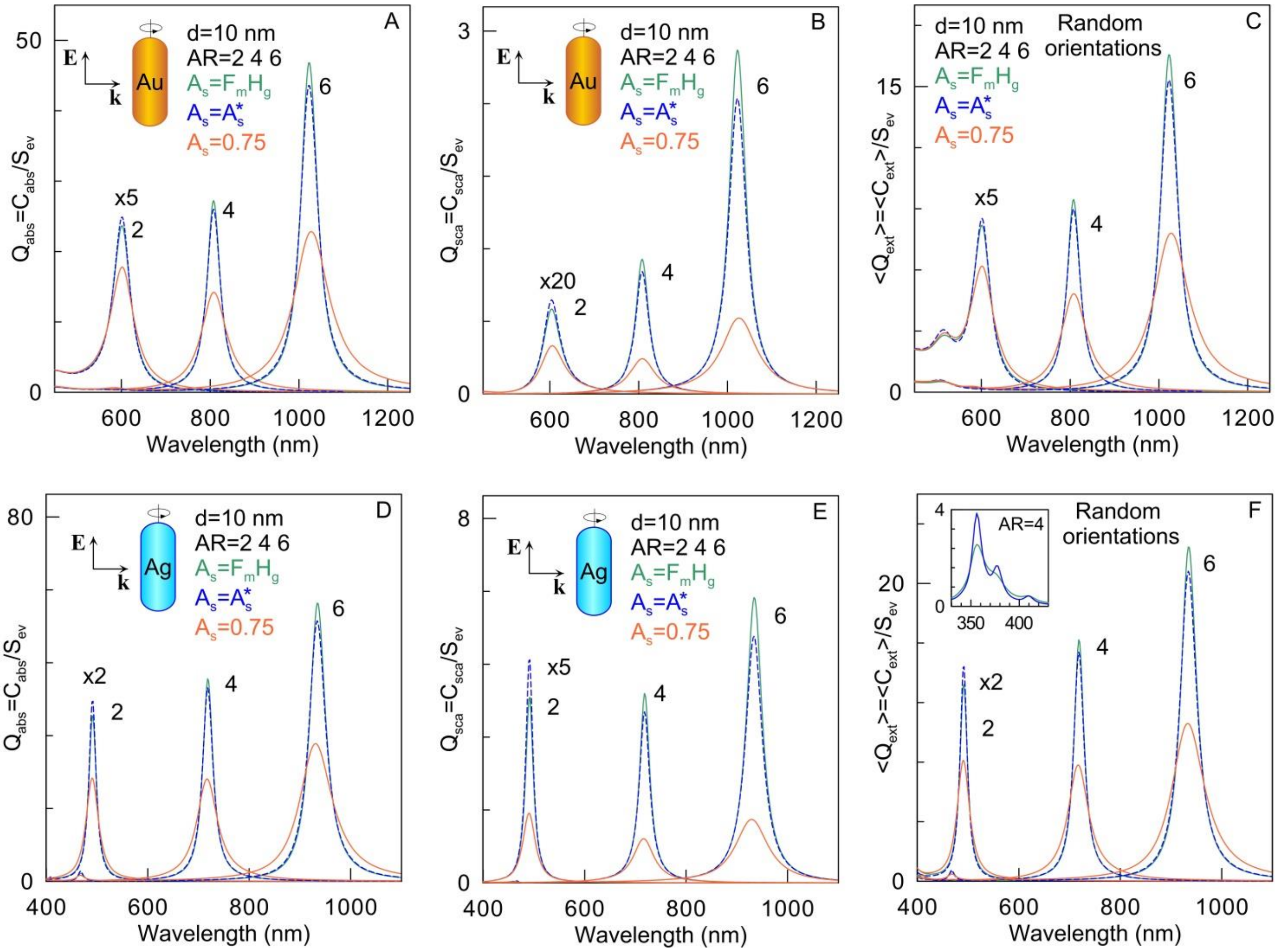

**Figure 3**. Absorption (A, D), scattering (B, E), and extinction (C, F) spectra of gold (A–C) and silver (D–F) nanorods calculated for longitudinal excitation (A, B, D, E) and random orientations (C, F). Green lines: COMSOL (full anisotropic correction, $A_c^{x,z} = F_m H_g^{x,z}$); blue and red lines: TMM (isotropic corrections with Eq. (24) and $A_s = 0.75$). Nanorod parameters are given in the panels. The inset in panel F displays the short-wavelength region of the extinction spectra calculated for randomly oriented silver nanorods with an aspect ratio of 4.

A close inspection of the spectra in Figure 3 reveals good agreement between the COMSOL spectra calculated with the QM anisotropic model and those calculated using the TMM with the isotropic size- and shape-dependent correction according to Eq. (24). Moreover, Figure S9 shows that the isotropic surface-scattering correction with a constant, shape-independent parameter $A_s = 0.25$ (TMM calculations) yields reasonable agreement with the full anisotropic correction (COMSOL calculations, $A_c^{x,z} = F_m H_g^{x,z}$). Notably, the fitting constant $A_s = 0.25$ was derived from single-particle sphere [57] and nanorod [58] measurements, in which contributions from inhomogeneous polydisperse size and shape distributions were excluded. Thus, we arrive at the following important conclusion: the convenient practice of neglecting electron scattering anisotropy is justified, provided that size-corrected dielectric functions are used together with reasonable isotropic electron scattering constants. It should be emphasized that this conclusion holds for the main dipole resonances only. For the short-wavelength multipole resonances, the full anisotropic model yields significant broadening compared with the isotropic approximation using a constant fitting parameter $A_s$ (see the inset in Figure 3F).

**B. Gold and silver nanodisks**

In oblate nanospheroids, the internal field is homogeneous, and the principal values of the geometrical tensor $H_g^{x,z}$ are calculated analytically (Eqs. (S12) and (S14)). Therefore, the spectral dependence of the scattering coefficients $A_s^{x,z}(\lambda)$ is determined solely by the material factor, according to Eqs. (13)–(15). However, for nanodisks, the principal values $A_s^{x,z}(\lambda)$ at a particular wavelength should be calculated iteratively, similar to those for nanorods. Figure 4 shows the calculated spectra of $A_s^{x,z}(\lambda)$ for gold and silver nanodisks. For comparison, the dashed lines show the scattering coefficients $(3/4)H_g^{x,z}$, in which the quantum mechanical material factor $F_m(\lambda)$ [47] has been replaced with a classical factor of 3/4 [7, 38, 51].

When the exciting field is directed along the major axis, perpendicular to the symmetry axis of silver disks, the spectral dependence of the principal component $A_s^x$ demonstrates a strong peak near 346–348 nm (Figure 4D). Similarly to silver nanorods, the peak positions do not depend on the disk aspect ratio of gold and silver nanorods.

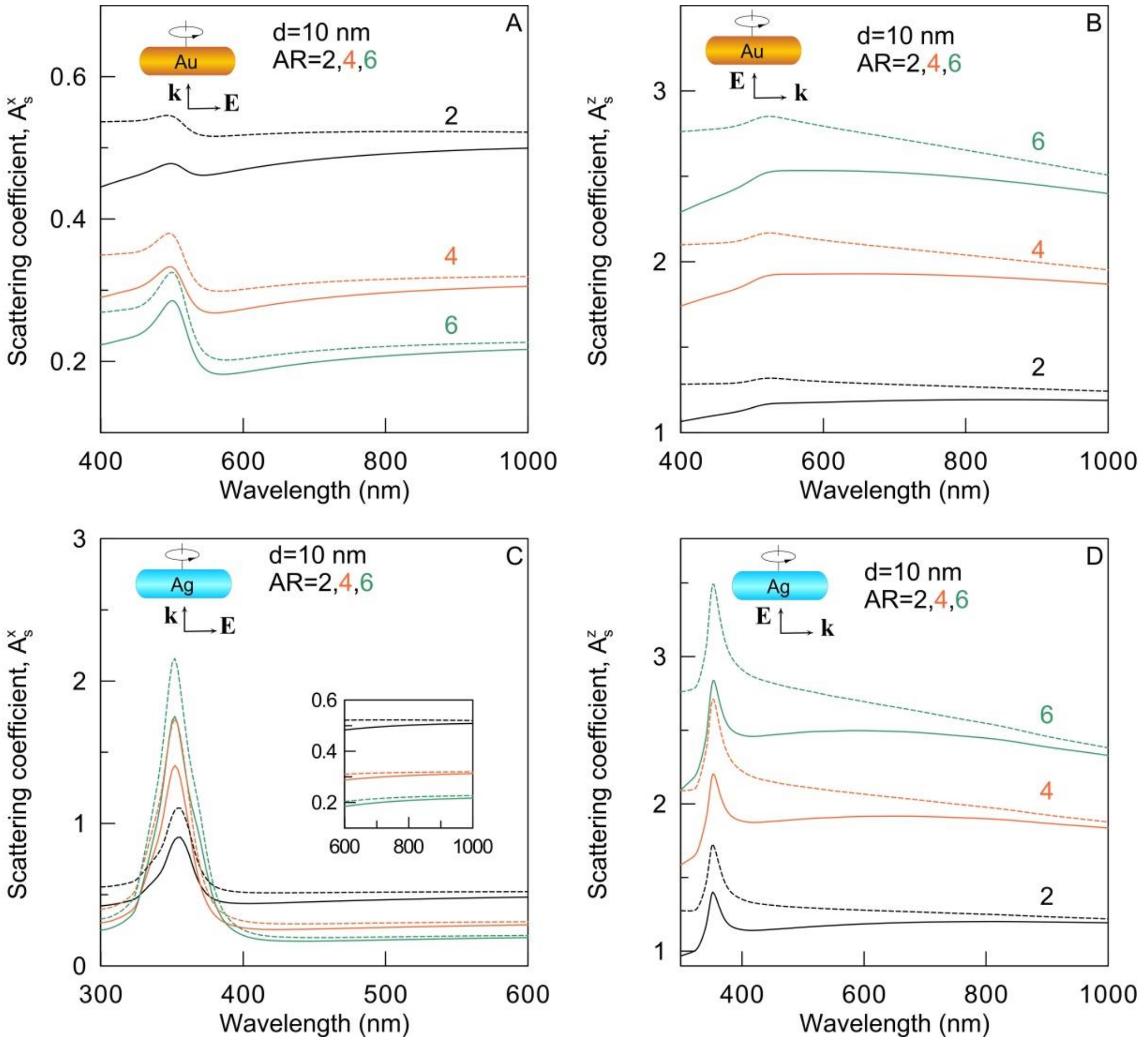

**Figure 4**. Spectra of the scattering coefficients $A_s^{x,z} = F_{mat} H_g^{x,z}$ (the principal values of the $A_s^{ij}$ tensor) for gold (A, B) and silver (C, D) nanodisks with a thickness of 10 nm and aspect ratios of 2 (black), 4 (orange), and 6 (green). For comparison, the dashed lines show the scattering coefficients $(3/4)H_g^{x,z}$ [38, 39]. Calculations were performed using COMSOL for ($\mathbf{E} \perp \mathbf{z}; \mathbf{k} \| \mathbf{z}$) and ($\mathbf{E} \| \mathbf{z}; \mathbf{k} \perp \mathbf{z}$) configurations, where $\mathbf{z}$ is the rotation axis. Data for the ($\mathbf{E} \perp \mathbf{z}; \mathbf{k} \perp \mathbf{z}$) configuration are close to those shown in panels A and C.

Figure 5 shows the absorption and scattering spectra of gold and silver disks calculated for excitation along the major axis, with the wave vector directed along the rotation axis z (configuration $\mathbf{E} \perp \mathbf{z}; \mathbf{k} \| \mathbf{z}$). Since the disks are small (thickness of 10 nm and maximum diameter of 60 nm), the absorption and scattering spectra for the ($\mathbf{E} \perp \mathbf{z}; \mathbf{k} \perp \mathbf{z}$) configuration are close to those shown in Figures 5A-E. Figures 5C and 5F show the extinction spectra of randomly oriented nanodisks. As in the case of nanorods, two models were used for comparison: one with an isotropic scattering constant $A_s = 0.75$ and one with $A_s^*(R_{ev}) = qA_s(d/2)$, where the scattering constant for the major semiaxis was set to 0.3 (see Eq. (23)). For nanodisks, the scaling parameter $q$ is given by equation

$$q(AR) = \frac{1}{AR}\left[1 + \frac{3}{2}(AR-1)\left(AR-1+\frac{\pi}{2}\right)\right]^{1/3}. \tag{25}$$

For the selected aspect ratios of 2, 4, and 6, the scaled scattering parameters are 0.25, 0.21, and 0.18, respectively. Remarkably, the first value coincides with the fitted scattering parameter derived from single-particle measurements [57, 58].

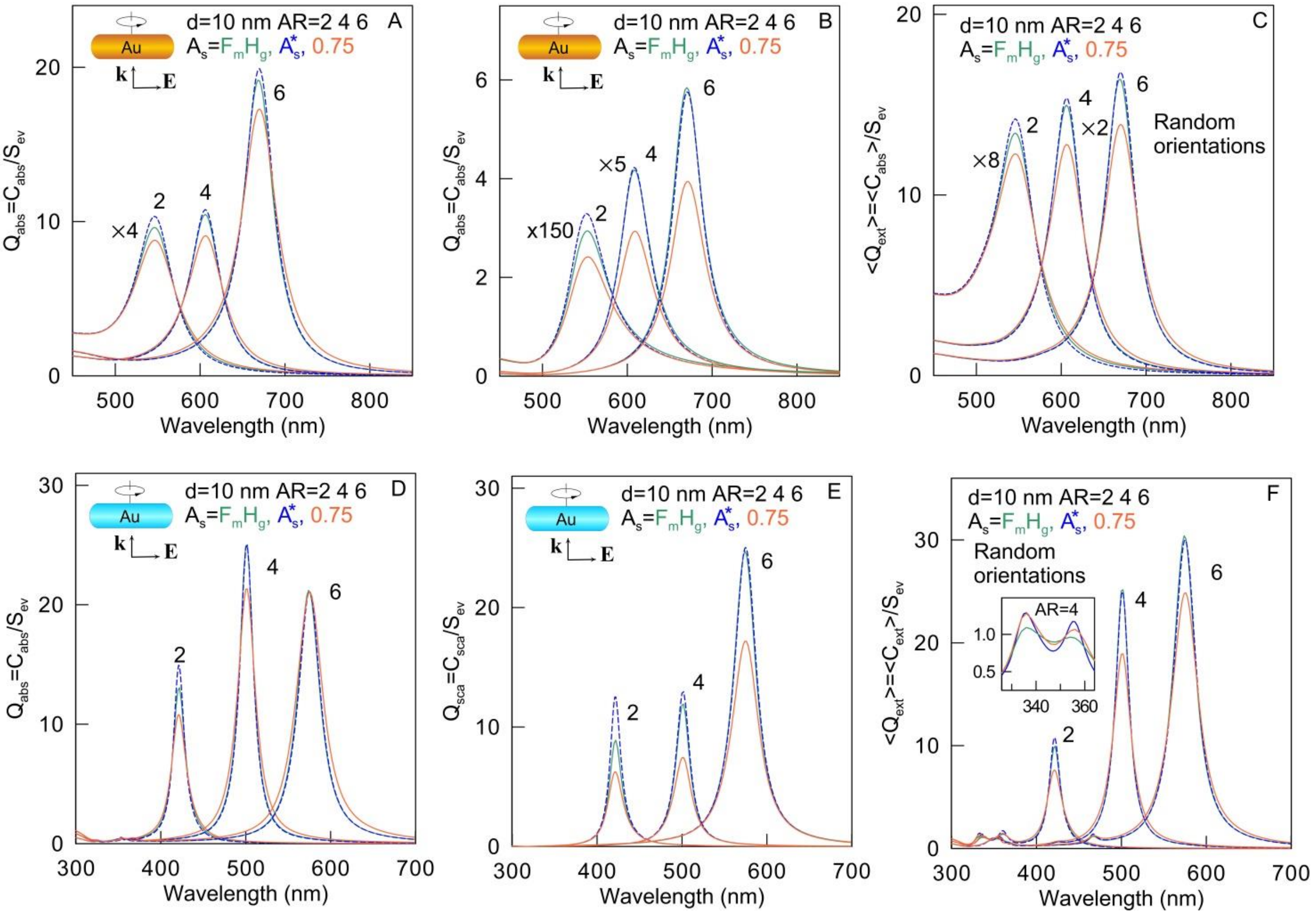

**Figure 5**. Absorption (A, D), scattering (B, E), and extinction (C, F) spectra of gold (A–C) and silver (D–F) nanodisks calculated for excitation (A, B, D, E) and random orientations (C, F). Green lines: COMSOL (full anisotropic correction, $A_c^{x,z} = F_m H_g^{x,z}$); blue and red lines: TMM (isotropic corrections with Eq. (25) and $A_s = 0.75$). Nanodisk parameters are given in the panels. The inset in panel F displays the short-wavelength region of the extinction spectra calculated for randomly oriented silver disks with an aspect ratio of 4.

From the spectra in Figure 5, we conclude that the absorption, scattering, and extinction spectra obtained with the full anisotropic correction can be accurately reproduced by an isotropic model employing an equivolume-radius scattering length and a scalar scattering constant, as given by Eq. (25). From a practical standpoint, an even simpler isotropic model with a scattering constant of 0.25 can be applied to nanodisk simulations. Again, we emphasize that this conclusion holds for the main dipole resonances only, whereas the full anisotropic model yields significant broadening of the short-wavelength multipole resonances (see the inset in Figure 5F).

**C. Physical mechanisms underlying spectral peaks in anisotropic scattering coefficients**

The existence of strong $A_s^{x,z}$ spectral peaks (Figures 4C and 4D) is somewhat counterintuitive, as they are observed when the exciting field is directed along the major axis, which is much greater than the mean free path of electrons. The same behavior has been obtained for silver nanorods under longitudinal excitation (Figure 1C), where the aspect ratio-independent peak position is located near 352-354 nm. To the best of our knowledge, the underlying physics of these unexpected peaks has not been discussed until now. Furthermore, this result is in strong contrast to calculations for geometrically similar oblate spheroids (Figure S10), where the internal field is homogeneous, and no peaks are observed. Consequently, the unexpected peaks in the spectra of $A_s^{x,z}$ can be attributed to an inhomogeneous distribution of the internal field within this spectral band. In this band, integration of the squared normal component of the internal field gives rise to a resonance-like feature. Notably, this resonance is distinct from the main dipole plasmons, which occur at longer wavelengths.

To give some insight into the underlying physical mechanism, we considered a detailed picture of extinction spectra of silver rods and disks in the 300–500 nm spectral band (Figure 6). For certainty, we chose a particular aspect ratio $AR = 4$ from our 2, 4, and 6 set and performed calculations for ten particular incidence angles from 0 to 90 degrees, two basic TE and TM polarizations, and for random orientations (analytical T-matrix solution). Figure 6 displays the calculated spectra for six excitation configurations ($TE_0$=$TM_0$, $TE_{45}$,$TM_{45}$, and $TE_{90}$,$TM_{90}$) of silver rods (A–C) and disks (D–F). The spectra for fixed and random orientations (not shown here) demonstrate specific short-wavelength peaks that essentially differ from well-known LSPR dipole and quadrupole (QD) excitations in axisymmetric plasmonic particles.

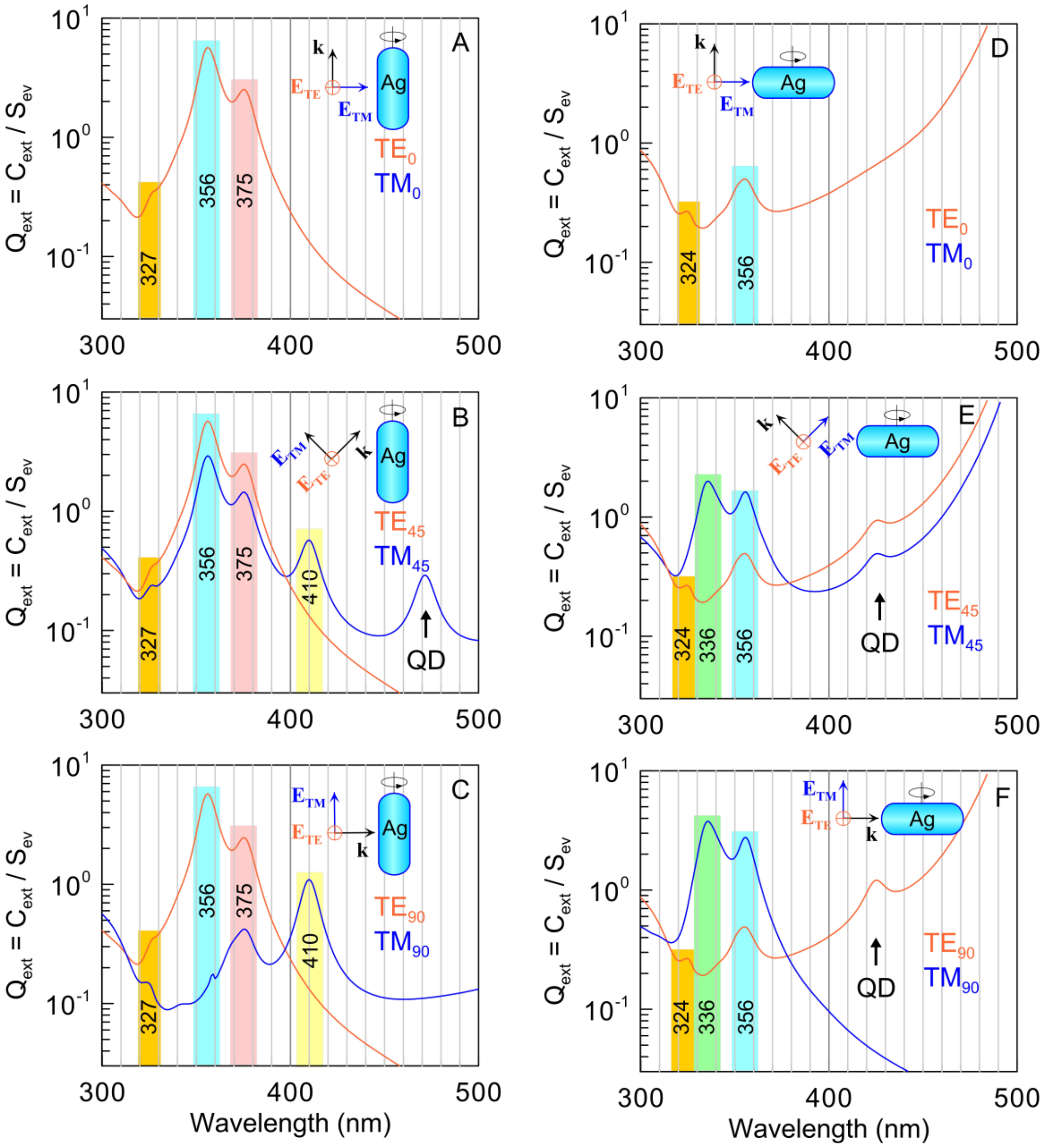

**Figure 6**. Extinction spectra of silver nanorods (A–C) and nanodisks (D–F) calculated by the T-matrix method for three incidence configurations ($TE_{0,45,90}$ –red, $TM_{0,45,90}$ – blue). The rod diameter

and disk thickness are 10 nm, the aspect ratio is 4, and the isotropic scattering constants $A_s^*$ were calculated using Eqs. (23) and (24) for rods and disks, respectively. The vertical arrows indicate quadrupole (QD) resonances; the dipole resonances are not shown. The vertical bars highlight specific plasmonic modes.

Since both the rod and the disk possess an axial symmetry and a plane of symmetry perpendicular to the axis of rotation ($z$), their geometric symmetry group is $D_{\infty,h}$, and all plasmon modes (resonances) can be classified according to the irreducible representations of this group [63]. For example, the quadrupole peaks at 472 nm for rods (symmetry $A_{1g}$) and at 425 nm for disks (symmetry $E_{2g}$) are excited if both the $\mathbf{E}$ and $\mathbf{k}$ vectors have nonzero projections on the major particle axis. For rods, the short-wavelength resonances at 327, 356, and 375 nm are excited under all configurations of the incident wave, whereas the resonance at 410 nm is observed only for TM configurations, except for incidence along the axis, when this mode is also not excited. For TM configurations under oblique or perpendicular incidence, the strongest dipole (718 nm) and quadrupole (472 nm) modes are excited in rods; these, unlike the UV resonances, are morphologically dependent on the shape factor. In contrast to nanorods, the spectrum of disks is dominated by strong two-dimensional in-plane modes with degeneracy in the plane. If the wave vector is directed along the symmetry axis of the disk, the electric vector in the disk plane is capable of exciting only the dipole mode, but not the quadrupole mode. A discussion of all the features of the short-wavelength resonances is beyond the scope of this work and will be presented elsewhere. For this section, the most important result of Figure 6 is the presence of a common plasmon peak near 356 nm, highlighted by the blue vertical bar. This peak is excited in both rods and disks and for all TE and TM configurations (except for TM90). Moreover, this peak is observed in a narrow range of 354–360 nm for nanorods and disks at all aspect ratios from 2 to 6.

If we now turn to the principal-value spectra of the electron scattering tensor in Figs. 1C and 4C, we observe the same universal peak near 360 nm for rods and disks with aspect ratios from 2 to

6. According to definitions (11) and (12), the resonant spectrum of the scattering tensor is determined by the resonant behavior of the geometric factor, since the material factor is a smooth function of frequency. In turn, the geometric factor is equal to the ratio of the surface-averaged squared modulus of the normal component of the field at the boundary inside the particle to the volume-averaged energy of the particle, which is proportional to the volume-averaged squared modulus of the internal field. To understand how these averaged quantities are physically formed for a rod and a disk, we calculated the field distribution inside and near the particle at the resonant wavelength of 356 nm, the distribution of the normal component of the field at the boundary inside the particle, and the distribution of charges at the boundary (Fig. 7).

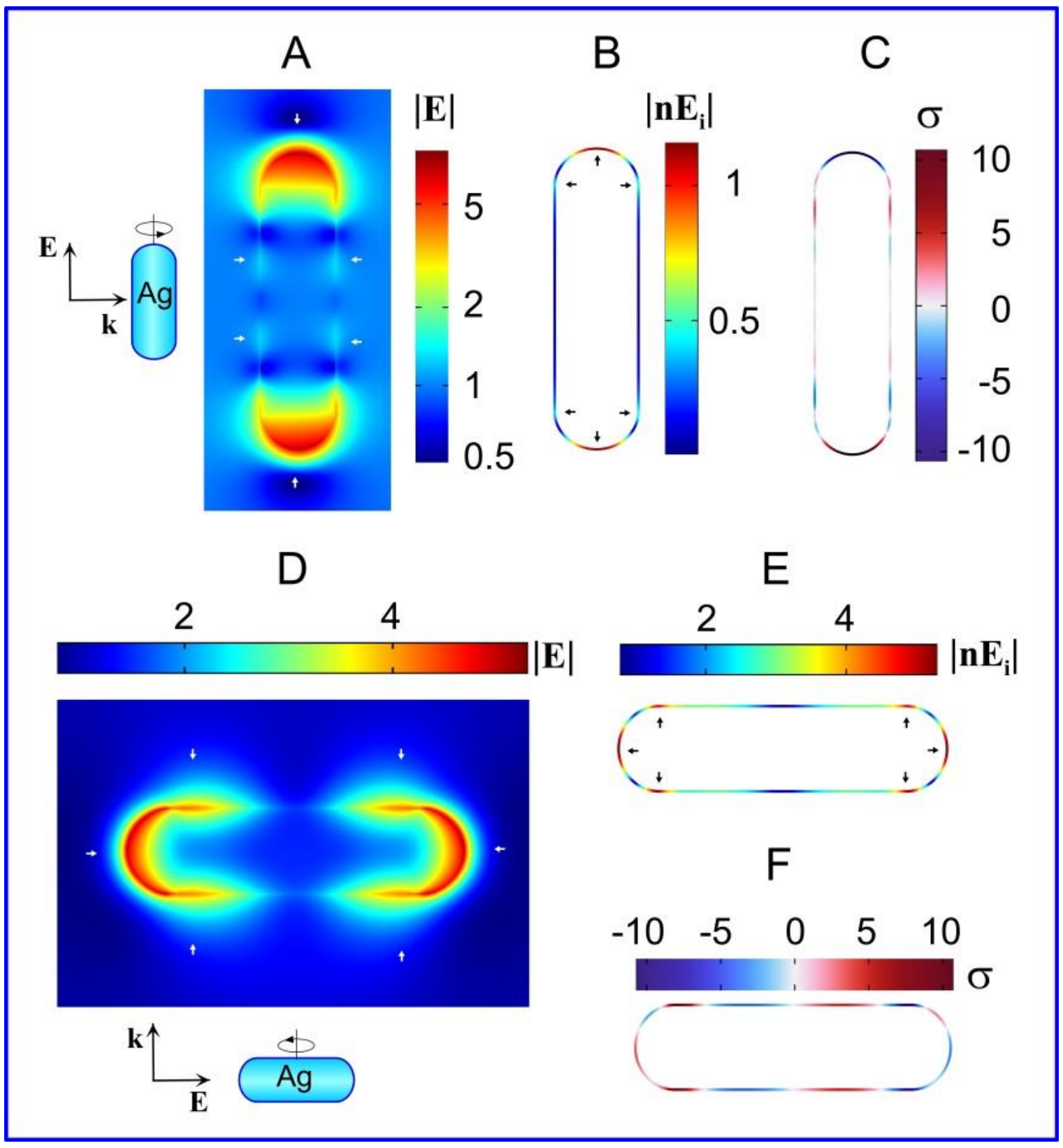

**Figure 7.** Distribution of the field modulus in the (x, z) cross-section inside and near a silver nanorod (A) and disk (D) under excitation along the long axis of the particle ($TM_{90}$ configuration for the rod and $TE_0$ for the disk) at the resonant wavelength of 356 nm. White arrows indicate the antinodes of plasmon excitations at the end faces and side surfaces of the rod, as well as at the edges and surfaces of the disk. Note that a logarithmic color scale is used in panel (A) to reveal the near-

field structure. Panels (B) and (E) show the distribution of the modulus of the normal component of the field. Black arrows indicate the maxima near the edges of the particles. Panels C and F show the distribution of the charge density on the surface of the particles.

Panels A and D show that the plasmon resonance at 356 nm is localized at the inner boundary of the hemispherical ends of the rod and the rounded edge of the disk. Unlike the conventional dipole resonance, the field practically does not extend beyond the particle boundary into the external dielectric medium. In both cases, the radius of curvature is only 5 nm, so that the field modulus at the inner boundaries is substantially greater than the exciting field. In the case of the nanorod, two further antinodes of this mode are observed, but their amplitude is much smaller than that at the end face. In the case of the disk, the in-plane field excites not only a localized surface plasmon at the rounded edge but also an additional intense mode on the flat surface of the disk (indicated by vertical white arrows).

The distribution of the normal component of the field at the inner boundary of the rods and disks reveals the presence of three regions on each edge: a central one and two lateral ones adjacent to the cylindrical or flat boundaries, respectively. Unlike the rod, the normal internal component of the field in the disk has a noticeable intensity along the entire diameter, with the exception of the central part. Thus, the rounded edge generates a normal component over the entire flat surface of the disk. In Figure 8, we present the spectra of the integral contributions to the surface-averaged normal component of the field and the volume-averaged inner field from the cylindrical flat (1) and rounded edge (2) parts (see also Section S8 and Figure S11 in the SI file):

$$F_S = \frac{1}{S}\left[\int_{S_1} |\mathbf{nE}_i|^2 \, \mathrm{d}S + \int_{S_2} |\mathbf{nE}_i|^2 \, \mathrm{d}S\right] \equiv F_S^1 + F_S^2 , \tag{26}$$

$$F_S = \frac{1}{V}\left[\int_{V_1} |\mathbf{E}_i|^2 \, \mathrm{d}V + \int_{V_2} |\mathbf{E}_i|^2 \, \mathrm{d}V\right] \equiv F_V^1 + F_V^2 . \tag{27}$$

First, we note a significant difference in spectra for TE0=TM0 (panels A-C) and TE90 excitations. In the TE90 case, the quadrupole peaks at 425 nm appear because both k and E vectors have components in the disk plane. Second, although the exciting field is directed parallel to the surface of the disk, the integral of the normal component over the flat surface is comparable to the contribution from the disk edge. Third, the volume integral contribution from the cylindrical part demonstrates short-wavelength resonances similar to that from the surface contribution, but its resonance near 354 nm is less high-Q and somewhat shifted toward longer wavelengths. Therefore, the ratio of the surface and volume integrals exhibits resonant behavior with a maximum at 356 nm (panels C and D). At TE90 excitation, the spectrum in panel D demonstrates an additional quadrupole peak near 425 nm. Finally, we note a significant difference between the short-wavelength spectra in Figure 6 and spectra in Figure 8 that demonstrate only one universal peak at 356 nm and the quadrupole peak at 425 nm.

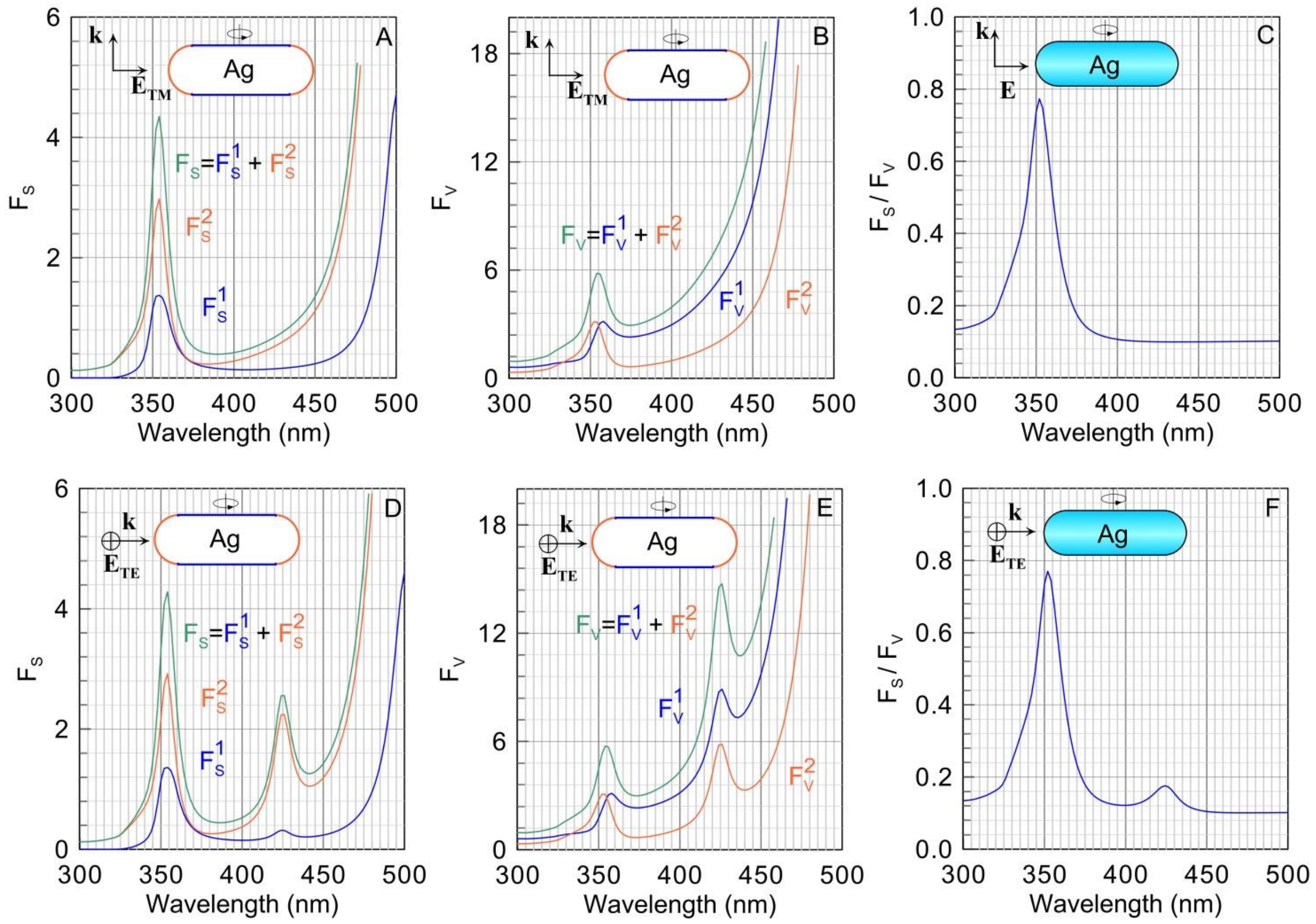

**Figure 8**. UV spectra of the integral contributions to the surface-averaged normal component of the inner field at the particle boundary (A, D) and to the volume-averaged inner field (B, E) for the cylindrical flat (superscript 1, blue lines) and rounded-edge (superscript 2, orange lines) parts under $TE_0 = TM_0$ (A-C) and $TE_{90}$ (D-F) excitations. Green lines represent the total summed quantities. Panels C and D show the UV spectrum of the total $F_S / F_V$ ratio, which is proportional to the $H_g^{\mathbf{e}}$ geometrical factor (see Eq. (12) in the main text). Calculations were performed for a silver disk with a thickness of 10 nm and aspect ratio of 4.

Based on the universal excitation of the 356 nm resonance under all incident-field configurations and for all aspect ratios of the rods and disks, as well as on the distributions of the local-field magnitude and the normal field component at the inner boundary of the rod and disk, we hypothesize that the physical origin of this resonance is associated with the excitation of a surface plasmon at the silver–water interface. As is well known, the condition for the excitation of surface plasmon polaritons at an interface between two media (in our case, silver–water) has the form $\varepsilon_{Ag}(\omega) = -\varepsilon_{H_2O}(\omega)$, or

$$\varepsilon_r(\omega) + 1 = 0, \qquad (28)$$

where $\varepsilon_r = \varepsilon_{Ag} / \varepsilon_{H_2O}$. If our assumption concerning the physical nature of the 356 nm resonance is correct, we expect a resonance of the material function $1/\left|\varepsilon_r(\lambda)+1\right|$ near 356 nm. In Figure S12, we present plots of the material function $1/\left|\varepsilon_r(\lambda)+1\right|$ for silver (panel A) and gold (panel B) in water. As can be seen, in accordance with the surface plasmon-polariton excitation condition (28), the resonance of this function for silver in water occurs at 352 nm, just near 356 nm. We believe that this coincidence is not accidental, since this resonance is determined solely by the properties of the dielectric function of silver in water and, therefore, its position does not depend on the details of the particle shape, in contrast to the conventional localized dipole and quadrupole resonances (LSPRs).

As is well known, these LSPR resonances are morphologically dependent on the size and, primarily, on the shape of the particles.

For gold disks in water, the surface scattering parameter $A_s^x$ also exhibits a weak (compared with silver) but likewise aspect-ratio-independent resonance near 510 nm (Figure 4A), in agreement with the position of the material-function resonance in Figure S12B. However, for gold nanorods, the resonance of the $A_s^x$ parameter is less pronounced and depends to some extent on the aspect ratio. This may be attributed to the difference in the position of the interband transition frequency for gold (517 nm) and silver (318 nm). Consequently, for silver, interband transitions do not affect the 356 nm resonance, whereas for gold, the 510 nm resonance overlaps with the interband transition edge.

**D. Gold and silver bicones**

Recently, we compared the extinction spectra, local refractive index sensitivity, SERS enhancement factors, and photothermal efficiencies of gold nanorods and gold bipyramids (AuBPs) for bacterial killing [64 ]. AuBPs exhibited a superior Q-factor in the UV–vis–NIR region, better analytical and SERS performance, and higher light-to-heat conversion efficiency. This superior plasmonic performance of AuBPs [65, 66] is attributed to differences in particle morphology: compared with gold nanorods, which have a cylindrical shape and hemispherical ends, AuBPs have a pentagonal cross-section [67] and a bicone-like overall shape with sharp tips. It has been shown [68] that the UV–vis spectra of experimental pentagonal bipyramids can be accurately modelled by bicones with properly defined geometrical parameters. This greatly simplifies plasmonic simulations because an equivalent bicone possesses axial symmetry, allowing the use of an efficient 2.5D approach [35] instead of time-consuming full 3D modelling.

Based on experimental data [68, 69], we used a constant bicone diameter of 20 nm and set the rounding radii at the bicone base and tips to 6 nm and 3 nm, respectively. These rounding radii are close to the experimental values but significantly greater than those used previously [70]. All simulations were performed for three aspect ratios, 2.5, 3.5, and 4.5, which correspond to NIR plasmon resonances in the 700–1000 nm range.

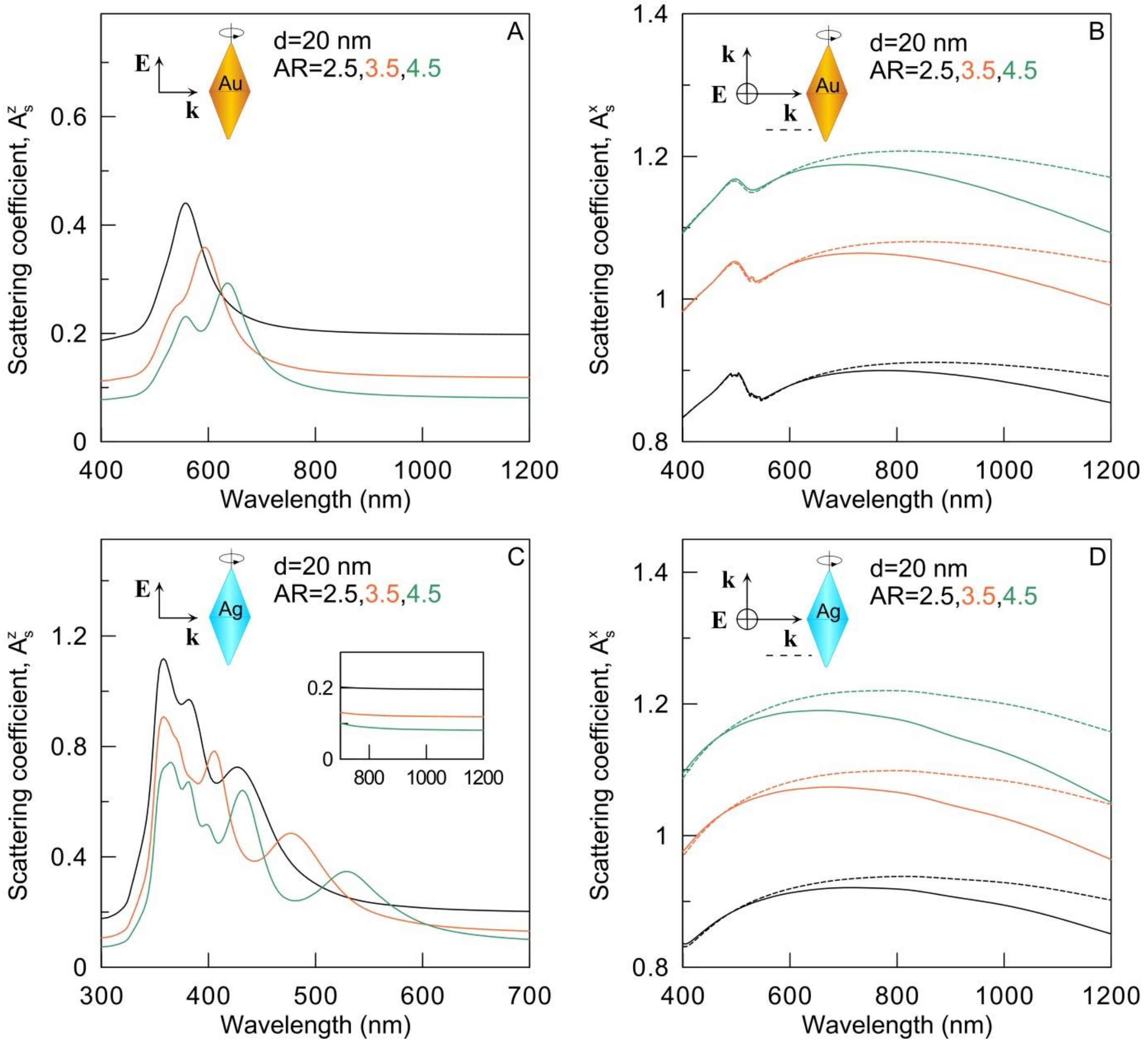

**Figure 9**. Spectra of the scattering coefficients $A_s^{z,x}$ (the principal values of the $A_s^{ij}$ tensor) for gold (A, B) and silver (C, D) bicones with diameters 20 nm and aspect ratios of 2.5 (black), 3.5 (orange), and 4.5 (green). In panels C and D, solid and dashed lines show calculations for the TE0 ( $\mathbf{E} \perp \mathbf{z}; \mathbf{k} \parallel \mathbf{z}$ ) and TE90 ( $\mathbf{E} \perp \mathbf{z}; \mathbf{k} \perp \mathbf{z}$ ) configurations, respectively.

Figure 9 shows the spectral dependence of the two principal components of the surface electron scattering tensor, $A_s^z$ (A, C) and $A_s^x$ (B, D), for gold (A, B) and silver (C, D) bicones. For gold bicones, the calculated spectra resemble those of gold nanorods (Figs. 1A, B). The key observation is that the perpendicular component $A_s^x$ is an order of magnitude larger than its parallel counterpart $A_s^z$. As with gold nanorods, the main short-wavelength peak of $A_s^z$ shifts to longer wavelengths with increasing aspect ratio; however, the peak itself splits into two components, and its left-hand component still exhibits a universal position near 560 nm. For silver bicones, the spectra of the longitudinal component $A_s^z$ exhibit a very complex multipole structure, especially at the highest aspect ratio of 4.5. Nevertheless, a universal peak near 360 nm is still observed, similar to that discussed in Section C. This remarkable universality is consistent with the discussion of its physical origin in Section C.

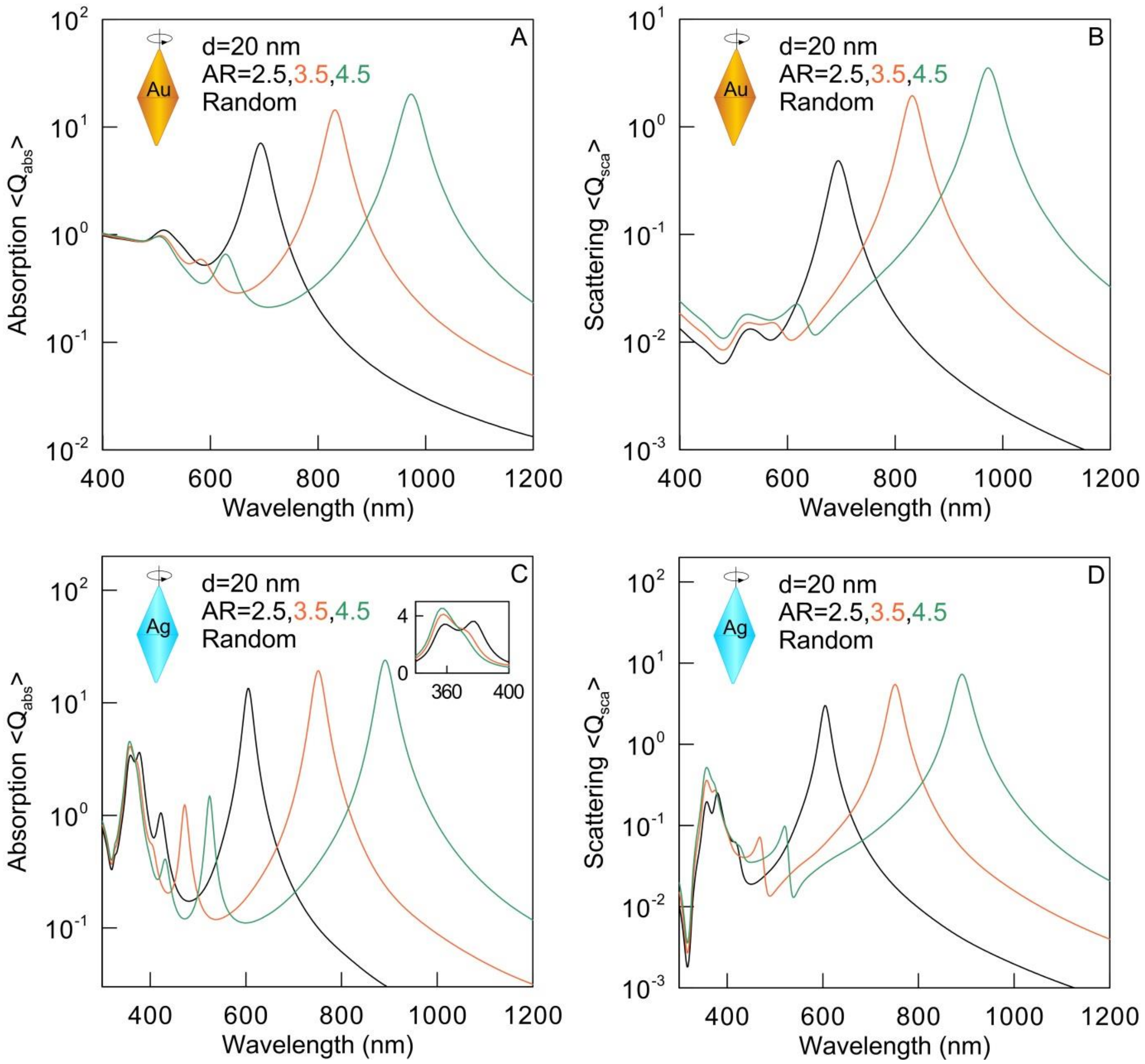

**Figure 10**. Absorption (A, C), scattering (B, D), spectra of gold (A, B) and silver (C, D) bicones calculated for random orientations with full anisotropic correction $A_c^{x,z} = F_m H_g^{x,z}$ (solid lines).

Bicone parameters are given in the panels. The inset in panel C displays a universal 360-nm peak position for all aspect ratios.

The absorption and scattering spectra of randomly oriented bicones are largely similar to those of nanorods. Along with the usual dominant dipole peak, which depends on the aspect ratio, quadrupole peaks are also observed. The inset in panel C for silver bicones shows a universal peak near 360 nm that is independent of the shape factor. The physical origin of this peak is explained in Section C.

**CONCLUSION**

We used a combination of quantum-mechanical and classical approaches to study the dependence of the principal components of the electron surface-scattering tensor in nonspherical gold and silver nanoparticles. The commonly accepted practice is to approximate these tensors by scalar constants, whose values vary widely for different types of nanoparticles and for the dielectric matrices in which they are embedded. However, our calculations revealed a strong dependence of the principal components of the scattering tensor on the wavelength of light, on the shape and material of the particles, and on their orientation relative to the incident linearly polarized light. We found an unusual universal maximum near 360 nm in the spectra of the principal components of the surface-scattering tensor for silver particles of various shapes, as well as a similar maximum in the absorption and scattering spectra. Moreover, this maximum coincided with the position of the maximum of the material function $1/\left|\varepsilon_r(\lambda)+1\right|$ of silver in water. A detailed analysis enabled us to relate these observations to the excitation of a surface plasmon polariton at this wavelength.

In real colloids, one usually deals with ensembles of randomly oriented particles. Therefore, the central question of our study was whether the tensor nature of the scattering constant and its strong spectral dependence can be ignored when modeling the spectra of real colloids. What errors are introduced by this commonly accepted approach? Our main result is as follows: the absorption and scattering spectra of real, randomly oriented ensembles of nonspherical particles can be accurately modeled using a reasonably chosen scalar approximation of the scattering constant and a scalar characteristic length, such as the equivalent-volume sphere radius or the Coronado–Schatz characteristic length $4V/S$. For particles that are monodisperse in size, the effective scattering parameter depends on particle shape and is smaller than conventional estimates around the classical value of 3/4. For example, it is closer to the experimental value of 0.25 when the equivalent-volume sphere radius is used as the characteristic length. This indicates that in colloids the main factor responsible for broadening is the polydispersity and polymorphism of the samples rather than plasmonic broadening effects caused by an increase in the damping constant.

We used a combination of quantum-mechanical and classical approaches to study the dependence of the principal components of the electron surface-scattering tensor in nonspherical gold and silver nanoparticles. The commonly accepted practice is to approximate these tensors by scalar constants, whose range of values varies widely for different types of nanoparticles and for the dielectric matrices in which they are embedded. However, our calculations revealed a strong dependence of the principal components of the scattering tensor on the wavelength of light, on the shape and material of particles, and on their orientation relative to the incident linearly polarized light. We found an unusual universal maximum near 360 nm in the spectra of the principal components of the surface-scattering tensor for silver particles of various shapes, as well as a similar maximum in the absorption and scattering spectra. Moreover, this maximum coincided with the position of the maximum of the material function $1/\left|\varepsilon_r(\lambda)+1\right|$ of silver in water. A detailed analysis enabled us to relate these observations to the excitation of a surface plasmon polariton at this wavelength.

In real colloids, one usually deals with ensembles of randomly oriented particles. Therefore, the central question of our study was whether the tensor nature of the scattering constant and its strong spectral dependence can be ignored when modeling the spectra of real colloids. What errors are introduced by this commonly accepted approach? Our main result is as follows: the absorption and scattering spectra of real randomly oriented ensembles of nonspherical particles can be accurately modeled using a reasonably chosen scalar approximation of the scattering constant and a scalar characteristic length, such as the equivalent-volume sphere radius or the Coronado–Schatz characteristic length $4V/S$. For particles that are monodisperse in size, the effective scattering parameter depends on particle shape and is smaller than conventional estimates of the order of the classical value of 3/4. For example, it is closer to the experimental value of 0.25 when the equivalent-volume sphere radius is used as the characteristic length. This indicates that in colloids the main factor responsible for broadening is the polydispersity and polymorphism of the samples rather than plasmonic broadening effects caused by an increase in the damping constant.

**Supplementary Material**. Section S1. Particle models. Figure S1.
Section S2. Material factors for gold and silver. Figure S2.
Section S3. The geometrical factor for spheroids. Figure S3.
Section S4. Convergence of iterations for the surface scattering constant. Figure S4
Section S5. Numerical orientation averaging of the light scattering cross sections. Figures S5-S7.
Section S6. Comparison of the surface scattering constants for nanorods and spheroids. Figure S8.
Section S7. Additional spectra for gold and silver nanorods. Figures S9, S10.
Section S8. Integral contributions to the surface-scattering broadening parameter from the flat and edge parts of the disk surface. Figure S11.
Section S9. Material functions $1/|\varepsilon_r + 1|$ for gold and silver in water. Figure S12
Section S10. The optical dielectric function of bulk silver. Figure S13

**Acknowledgments**

This research was supported by the state assignment of the Ministry of Science and Higher Education of the Russian Federation for the Federal Research Center "Saratov Scientific Center of the Russian Academy of Sciences,".

**AUTHOR DECLARATIONS**

**Conflict of Interest**

The authors have no conflicts to disclose.

**Author Contributions**

Nikolai Khlebtsov: Conceptualization; Formal analysis; Investigation; Writing; Supervision

Sergey Zarkov: COMSOL simulations; Data analysis

**DATA AVAILABILITY**

The data supporting the findings of this study are available within the article and its supplementary materials. The T-matrix codes are freely available at https://github.com/khlebtsov/T- matrix- abs-host.

# Supplementary Material

# Origins of Optical Anisotropy in Nonspherical Plasmonic Particles: The Role of Surface Electron Scattering

Nikolai G. Khlebtsov[1*], Sergey V. Zarkov[1,2]

[1]Institute of Biochemistry and Physiology of Plants and Microorganisms, "Saratov Scientific Centre of the Russian Academy of Sciences," 13 Prospekt Entuziastov, Saratov 410049, Russia

[2]Institute of Precision Mechanics and Control, "Saratov Scientific Centre of the Russian Academy of Sciences," 24 Ulitsa Rabochaya, Saratov 410028, Russia

[*]To whom correspondence should be addressed. E-mail: (NGK) khlebtsov@ibppm.ru

CONTENT

## Section S1. Particle models

Figure 1 shows three NP models that are described by a characteristic size $L$, a generalized aspect ratio $AR$, and the complex metal permittivity $\varepsilon_1$.

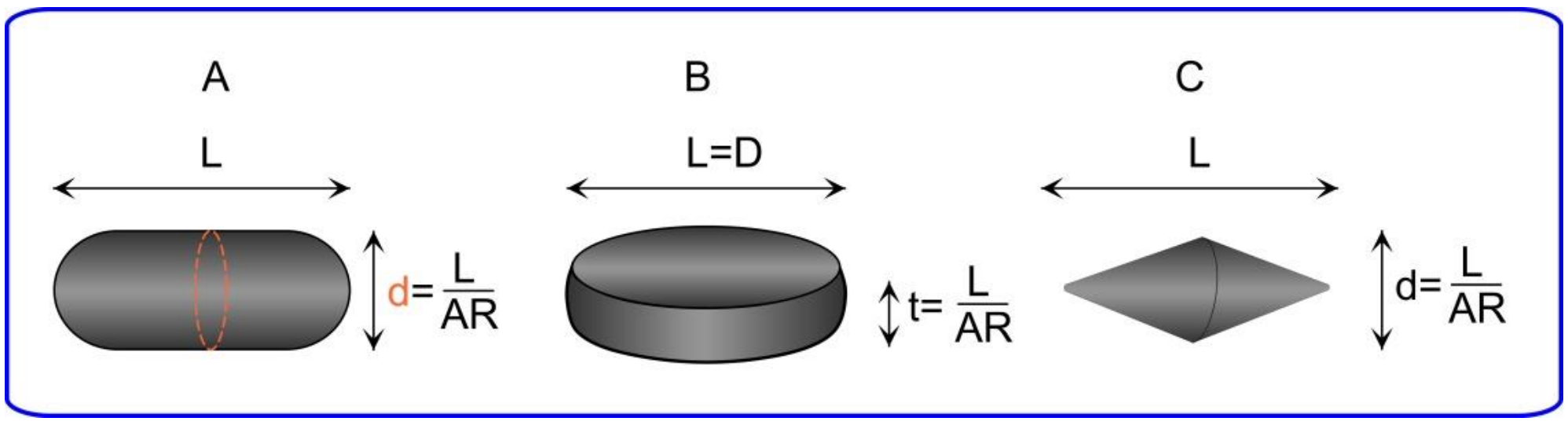

**Figure S1**. Geometrical models for the three particle morphologies, characterized by characteristic length $L$ and generalized aspect ratio $AR$. (A) Nanorods with hemispherical caps at the tips. (B) Nanodisks with a semicircular edge profile. For nanorods and nanodisks, the diameter is 10 nm, and the aspect ratios are 2, 4, and 6. Bicones have a diameter of 20 nm, aspect ratios of 2.5, 3.5, and 4.5, and rounded tips and circular base edges with rounding radii of 3 nm and 6 nm, respectively.

## Section S2. Material factors for gold and silver

We define the material factor $F_{mat}(\omega)$ by the following expressions

$$F_{mat}(\omega) = C_{mat} G(x = \hbar\,\omega / \varepsilon_F\,), \quad C_{mat} = \frac{4\alpha}{\pi}\frac{c}{v_F}\left(\frac{\varepsilon_F}{\hbar\omega_p}\right)^2, \tag{S4}$$

where our function

$$G(x) = \frac{2}{x}\int_0^1 du(1-u)\sqrt{u(u+x)} - \frac{2}{x}\int_0^{1-x} du\,(1-x-u)\sqrt{u(u+x)}\,, \tag{S5}$$

differs from that of Ref.[1] by normalization ($G(0)=1$). The integrals (S5) can be taken analytically by applying the substitution $u = t^2$ and using standard integrals [2] (232.01 and 234.01)

$$\int\sqrt{u(u+x)}du = \int 2t^2\sqrt{t^2+x}dt = 2\left[\frac{tr^3}{4} - \frac{xtr}{8} - \frac{x^2}{8}\ln(t+r)\right], \quad r = \sqrt{t^2+x} \tag{S6}$$

$$\int u\sqrt{u(u+x)}du = \int 2t^4\sqrt{t^2+x}dt = 2\left[\frac{t^3r^3}{6} - \frac{xtr^3}{8} + \frac{x^2tr}{16} + \frac{x^3}{16}\ln(t+r)\right], \quad r = \sqrt{t^2+x} \tag{S7}$$

After substituting these integrals into (S5) and some tedious algebra, we arrive at the following analytical solution

$$G(x) = \frac{2+3x}{6x}r^3 - \frac{2+x}{4}\left[r + xp\right] - \frac{(1-x)}{2}\left[\frac{y(2-x)}{x} - xq\right] + \frac{2y^3}{3x} - \frac{y(2-x)}{4} + \frac{qx^2}{4}\,, \tag{S8}$$

$$r = \sqrt{1+x}\,,\ y = \sqrt{1-x}\,,\ p = \ln\left((1+r)/\sqrt{x}\right),\ q = \ln\left((1+y)/\sqrt{x}\right), \tag{S9}$$

which coincide with Eq. (15) in the main text.

For low photon energy, Uskov et al. [3] also provided another expression for the material factor with function $G_1(x)$ given by Eq. (16) of the main text. For numerical estimates in the case of gold, we will employ the values cited in our article [4], which are practically identical to those reported in [1]:

$$\hbar\omega_p = 9.0 \pm 0.4 \text{ eV},\ E_F = 5.51 \text{ eV},\ \text{v}_F = 1.4\times10^8 \text{ см/с},\ \gamma_b = 1.06\times10^{14}\text{s}^{-1}. \tag{S6}$$

In Ref. [5], a very close value of $\hbar\omega_p = 9.03$ eV is given. Figure S2A shows functions $G(x)$ as given by Eq. (15) and (16) of the main text. The material factors $F_{mat}$ for gold and silver in panel B differ slightly because of a small difference in $C_{mat}$ constants: 0.7464 (Au) and 0.7632 (Ag).

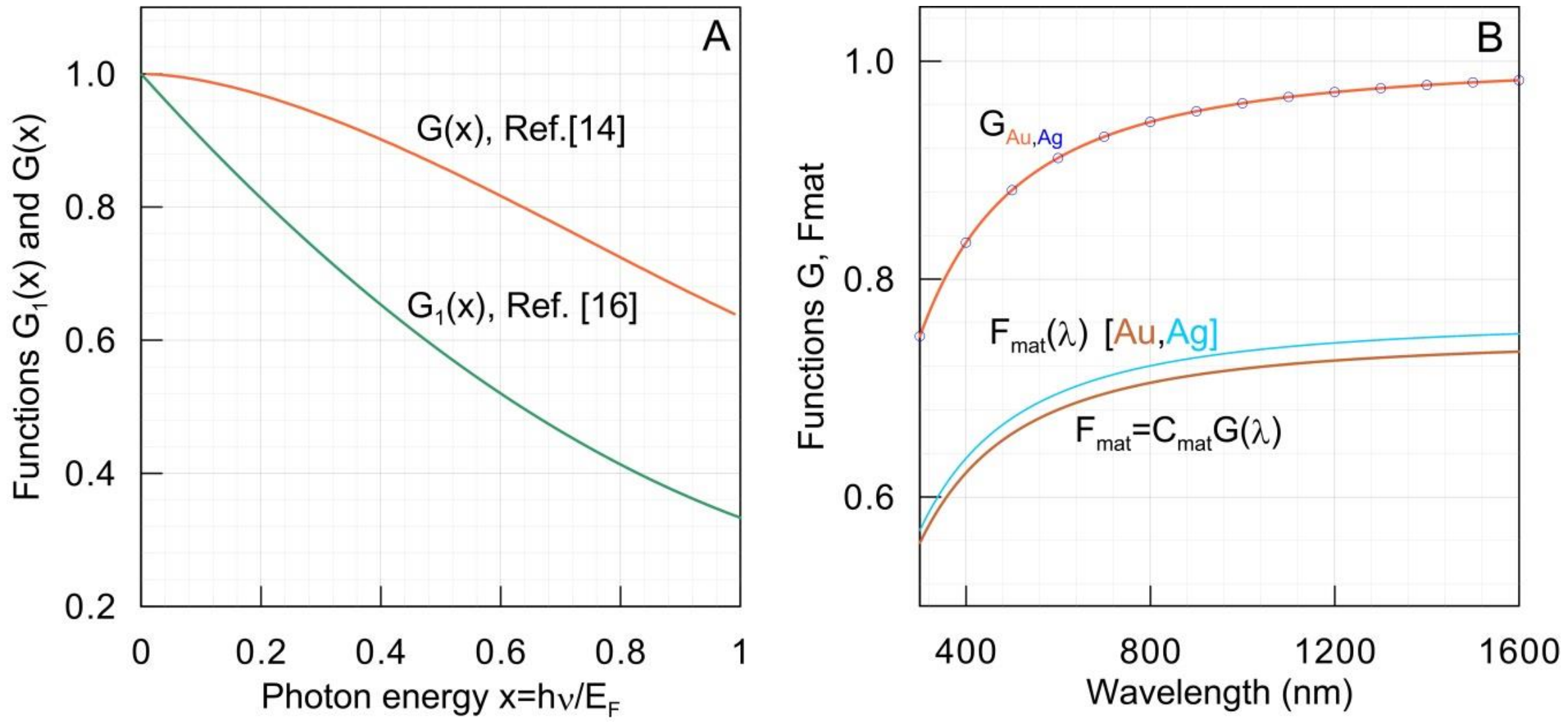

**Figure S2**. Functions $G(x), G_1(x), \quad x = \hbar\nu / E_F$ (A) and $G(\lambda)$, $F_{mat}(\lambda) = C_{mat}G(\lambda)$ (B) for gold and silver. Functions $G(\lambda)$ for gold and silver are practically identical (the orange solid line and blue circles in panel B). The material factors $F_{mat}$ for gold and silver are slightly different because of a small difference in $C_{mat}$ constants (Eq. (S4)): 0.7462 (Au) and 0.7632 (Ag).

The plasma frequency, Fermi energy and Fermi velocity of silver [6, 7] are practically indistinguishable from those of gold:

$$\hbar\omega_p = 8.9 \pm 0.2 \text{ eV}, \ E_F = 5.49 \text{ eV}, \ \mathrm{v}_F = 1.39 \times 10^8 \text{ cm/s}. \qquad \text{(S7)}$$

In Ref. [8], the value $\hbar\omega_p = 8.7$ eV was obtained, which falls within the scatter range of (S7). Consequently, we arrive at essentially the same $G_i(x)$ function and the material factor $F_{mat}(\lambda)$ for silver (Figure S2).

The reported bulk damping constant of silver varies in a broad range: $\tau_b(\text{fs}) = 1/\gamma_b = 36$ [7], 8.6 [9], 15-23 [10], 14.5 [7], 14.3 [11] (the latter two values refer to polycrystalline samples); for single crystals, Otter [12] obtained a much larger value of 31.3 fs. This disparity is likely attributable to the fact that the relaxation time is greater for single crystals. Jiang et al. [13] obtained an approximately constant value of 15 fs, decreasing to 5 fs in the 800–400 nm range. Thus, a value of 15 fs represents a reasonable estimate for polycrystalline samples, but may be on the order of 30 fs for single crystals. Given the uncertainty in the values of the parameters employed, one may, for silver, use the material factor for gold (Figure S2) and vice versa.

## Section S3. The geometrical factor for spheroids

We consider a particle with a rotation axis $c$ directed along the $z$ axis of the coordinate system $(x, y, z)$ associated with the particle. For spheroids, rods, and disks, the characteristic semiaxes are given by $(a, b = a, c)$, where for a rod the length and diameter are $L = 2c$ and $d = 2a$, respectively; similarly, for a disk the thickness and diameter are $t = L = 2c$ and $D = 2a$. We define the shape factor as the ratio of the larger dimension to the smaller one, such that $AR \equiv R = c / a = L / d \geq 1$ for prolate particles and $AR \equiv R = a / c = D / t \geq 1$ for oblate particles.

The electric field is assumed to be directed either along the symmetry axis $c$ ($\| \equiv z$) or perpendicular to it ($\perp \equiv x$). For spheroids, the internal field is isotropic and parallel to the external field applied along the symmetry axis; therefore,

$$H_g^{\mathbf{e}} = \frac{a_0 \int (\hat{\mathbf{n}}\mathbf{E}_i)^2 \mathrm{d}S}{\int |\mathbf{E}_i|^2 \,\mathrm{dV}} = \frac{\frac{a_0}{V} \int (\hat{\mathbf{n}}\mathbf{E}_0)^2 \mathrm{d}S}{\frac{1}{V} \int |\mathbf{E}_0|^2 \,\mathrm{dV}} = \frac{\frac{a_0}{V} \int (\hat{\mathbf{n}}\mathbf{E}_0)^2 \mathrm{d}S}{E_0^2} = \frac{3}{4\pi a_0^2} \int_S (\hat{\mathbf{n}}\hat{\mathbf{e}}_0)^2 \mathrm{d}S , \tag{S11}$$

where $\hat{\mathbf{e}}_0 = \mathbf{E}_0 / E_0$. Consider first the case where the field vector is directed along the spheroid's symmetry axis, i.e., along the $z$ axis. Evaluation of the integral yields

$$H_g^z = \frac{3R^{2/3}}{2e^2}\left[1 - R^{-2}\mathrm{at}(e)\right] = \frac{3R^{2/3}}{2R^2 - 2}\left[R^2 - \mathrm{at}(R)\right], \tag{S12}$$

where we have introduced the function

$$\mathrm{at}(R) = \begin{cases} \dfrac{\mathrm{artanh}(e)}{e}, & e = \sqrt{1 - R^{-2}}, \\ \dfrac{\mathrm{arctg}(e_1)}{e_1}, & e_1 = \sqrt{R^2 - 1}, \end{cases} . \tag{S13}$$

For spheres $R = 1$, $\mathrm{at}(1) = 1$ and $H_g^z(1) = 1$. We note that formula (S12) can be reduced to the form given in equation (37) of Ref. [1]. For the case of field orientation transverse to the symmetry axis (along the x-axis), the formula for the geometric polarization factor can be written as

$$H_g^x = \frac{3R^{2/3}}{4(R^2-1)}\left[\text{at}(e)\frac{2R^2-1}{R^2}-1\right]. \qquad \text{(S14)}$$

For spheres, $H_g^x(1)=1$. The parameters $H_g^{x,z}$ depend on the polarization (as indicated by the superscripts) and on the particle shape. Figure S3 presents the results of calculating the parameters as a function of the axial ratio for prolate and oblate particles using analytical formulas. These results are in agreement with [1].

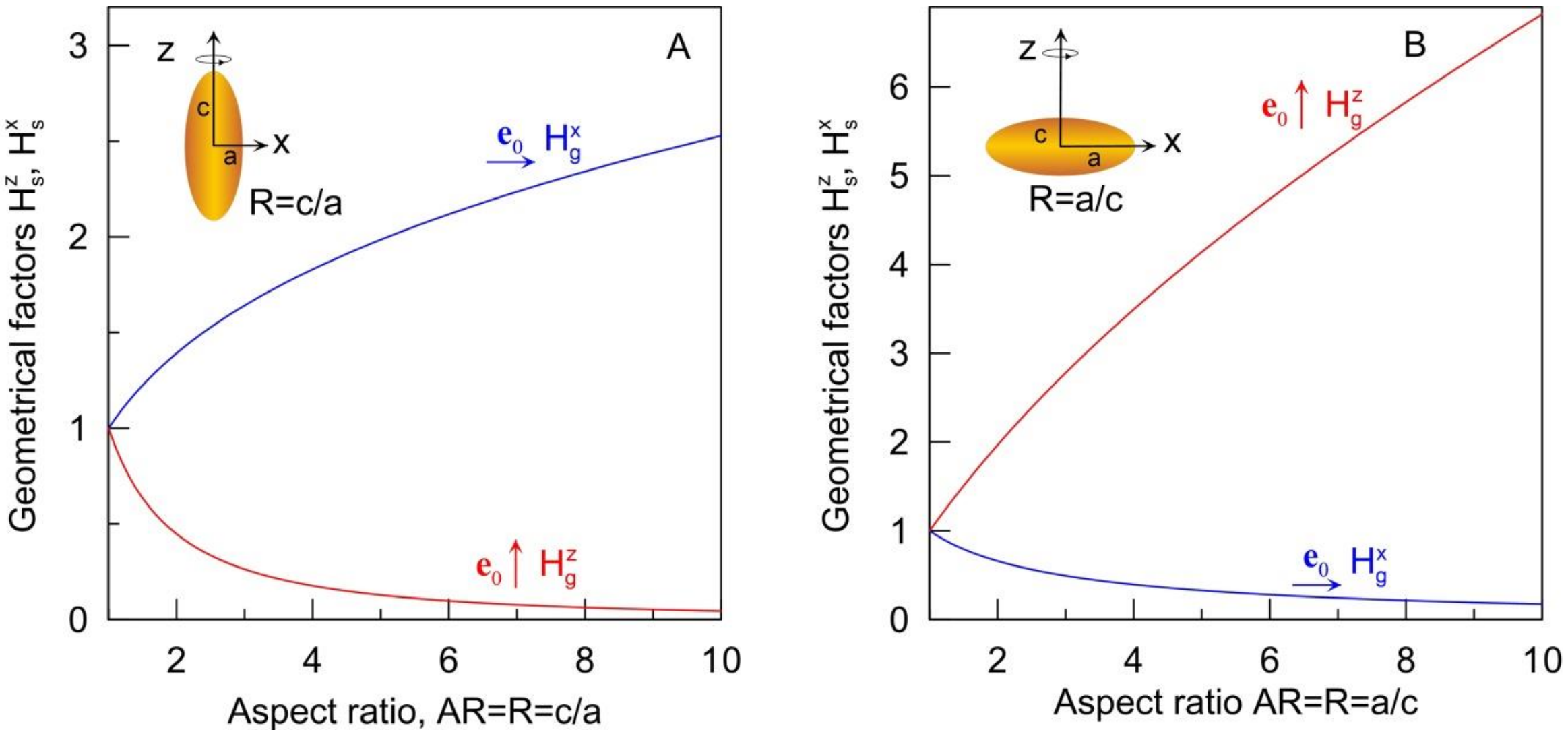

**Figure S3**. Dependence of the geometrical factors $H_g^x$ and $H_g^z$ on the aspect ratio of prolate (A) and oblate (B) spheroids.

As is evident from the plots, the parameters $H_g^{x,z}$ exhibit a sharp decrease when the field vector is oriented along the long axis of the spheroid. Conversely, for field orientation along the minor axis, the parameters $H_g^{x,z}$ increase to values of 2.5 in the case of prolate particles and up to 6.5 for thin disks. As expected, the contribution of anisotropic surface electron scattering becomes appreciable already at an axial ratio of about 4. It should also be noted that, according to the theory under consideration, for thin rods and disks the component of the dielectric function along the largest dimension of the nanoparticle remains virtually unperturbed by the electron scattering contribution and should be close to the values characteristic of the bulk metal.

If the external field is oriented at an arbitrary angle $\theta$ with respect to the symmetry axis of a small spheroid (TM configuration in Figure S5), then we can decompose the field vector of the incident field into two components, parallel and perpendicular to the symmetry axis

$$\mathbf{E}_0 = \mathbf{E}_{0x} \sin\theta + \mathbf{E}_{0z} \cos\theta \,. \qquad \text{(S15)}$$

By virtue of the superposition principle, the internal field remains homogeneous, just as in the case where the external field is aligned along either of the symmetry axes [14]. However, it will not be parallel to the external field owing to the anisotropy of the depolarization factors. The internal field is given by the linear relation

$$\mathbf{E}_i = \hat{p}\mathbf{E}_0 \,, \qquad \text{(S16)}$$

where the axial tensor $\hat{p}$ has two principal components along $p_z$ and perpendicular $p_x$ to the symmetry axis

$$p_{x,z} = \frac{1}{1 + L_{x,z}\left(\varepsilon_r - 1\right)} \,, \qquad \text{(S17)}$$

$\varepsilon_r = \varepsilon / \varepsilon_m$ is the complex relative permittivity of the metal, and $L_{x,z}$ are the well-known geometrical depolarization factors of spheroids [14]

$$L_z = \begin{cases} \dfrac{1-e^2}{2e^3}\left(\ln\dfrac{1+e}{1-e} - 2e\right), e = \sqrt{1 - AR^{-2}}, c \geq a, \\ \dfrac{1+e^2}{e^2}\left(1 - \dfrac{\mathrm{arctg}(e)}{e}\right) - \dfrac{2}{e^2}, e = \sqrt{AR^{-2} - 1}, c \leq a, \end{cases} \;, \; L_x = \frac{1 - L_z}{2} \,. \qquad \text{(S18)}$$

In Eq. (S18), the aspect ratio $AR = c / a \geq 1$ for prolate spheroids $(a, a, c)$ and $AR = a / c \geq 1$ for oblate ones. By inserting Eqs. (S16) into Eq. (S11), we obtain in the numerator two quadratic terms proportional to $\left(\mathbf{nE}_{ox}\right)^2$ and $\left(\mathbf{nE}_{oz}\right)^2$, as well as the cross-term $(\mathbf{nE}_{0x})(\mathbf{nE}_{0z})$

$$\int (\mathbf{nE}_i)^2 \mathrm{d}S = \int \left[ p_x^2 \sin^2\theta (\mathbf{nE}_{0x})^2 + p_z^2 \cos^2\theta (\mathbf{nE}_{0z})^2 + p_x p_z \sin 2\theta (\mathbf{nE}_{0x})(\mathbf{nE}_{0z}) \right] \mathrm{d}S \,. \qquad \text{(S19)}$$

The quadratic terms are proportional to the main components $H_g^{x,z}$ as derived above, whereas the cross-term vanishes upon integration over the spheroid surface. Similarly, the cross-terms vanish upon integration over the particle volume. Thus, we arrive at the following solution for the TM incident field

$$H_g^e(\theta) = \frac{H_g^x p_x^2 \sin^2\theta + H_g^z p_z^2 \cos^2\theta}{p_x^2 \sin^2\theta + p_z^2 \cos^2\theta} \,. \qquad \text{(S20)}$$

After replacing $\theta = \pi / 2 - \vartheta$ , we arrive at Eq. (20) of the main text. If the incident TE field is perpendicular to the plane $(\mathbf{k}, \mathbf{c})$, the geometrical factor does not depend on the incidence angle $\vartheta$ (Figure S5) in the electrostatic approximation.

## Section S4. Convergence of iterations for the surface scattering constant

The iteration scheme for calculating the surface scattering constant is similar to that described previously [15] for the surface damping parameter. As the material factor in Eqs. (11) and (14) of the main text do not depend on the particle dielectric function, we have to calculate the geometrical factors by iteration only. For the initial zero iteration, we used the analytical geometrical factors for spheroids (Eqs. (S12) and (S14)). Figure S4 shows the dependence of the percentage error on the iteration number.

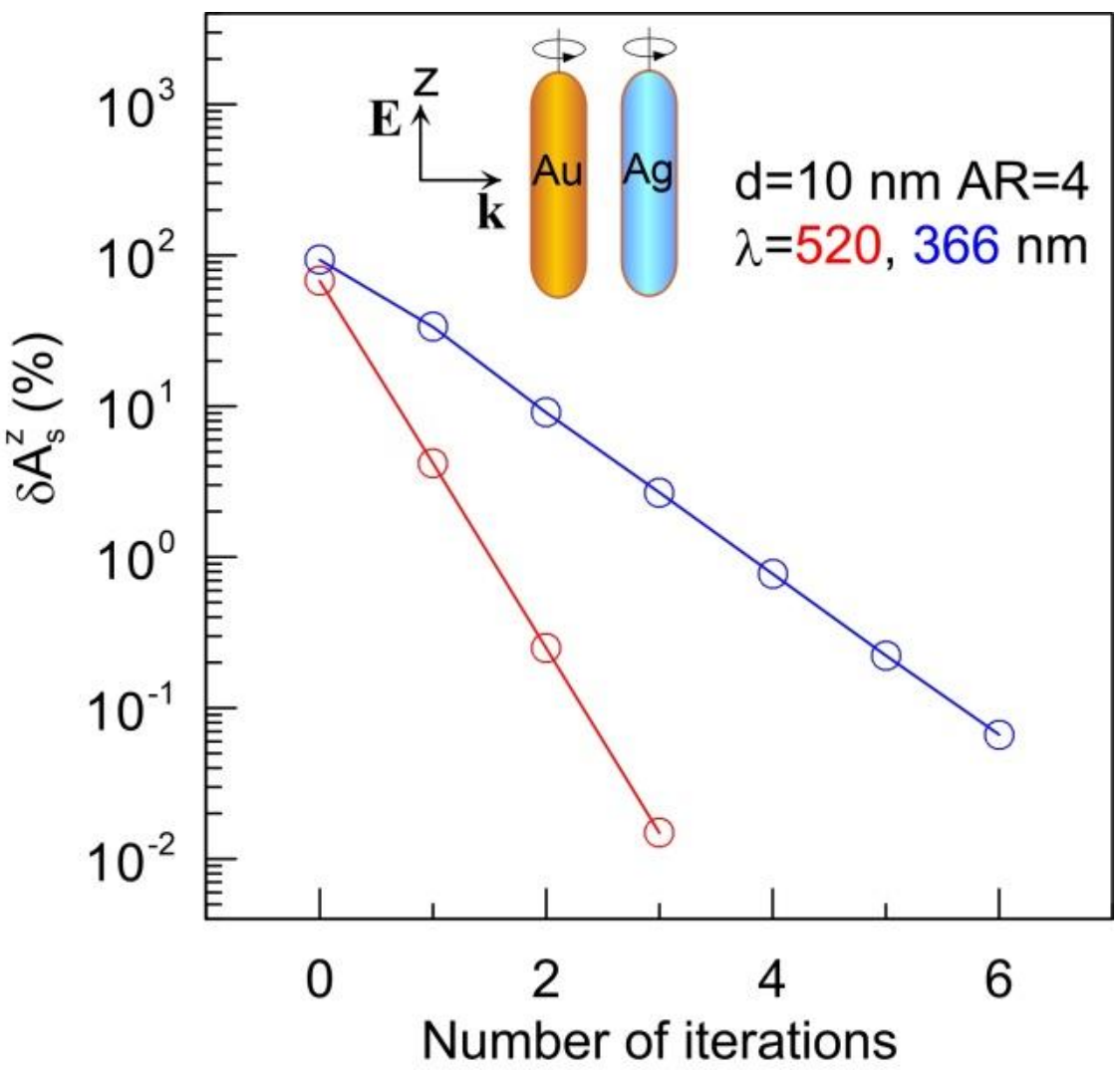

**Figure S4**. Dependence of the relative error $\delta A_s^z(\%)$ on the number of iterations in iterative calculations of the surface scattering constant $A_s^z$ for gold and silver nanorods. For an aspect ratio of 4, at wavelengths of 520 nm (gold) and 370 nm (silver), the errors are maximal, and convergence is slowest. In practical simulations, two iterations are sufficient for gold and four for silver.

## Section S5. Numerical orientation averaging of the light scattering cross sections

The most effective and accurate approach to calculate the orientation-averaged extinction and scattering cross sections is given by the T-matrix equations [16, 17]

$$\langle Q_{ext} \rangle = \frac{\langle C_{ext} \rangle}{\pi R_{ev}^2} = -\frac{2}{k^2 R_{ev}^2} \mathrm{Re}\left[\mathrm{Tr}(\hat{T})\right] = -\frac{2}{k^2 R_{ev}^2} \mathrm{Re} \sum_{i=1}^{2} \sum_{n=1}^{\infty} \sum_{m=-n}^{n} T_{mn,mn}^{ii} . \quad \text{(S21)}$$

$$\langle Q_{sca} \rangle = \frac{\langle C_{sca} \rangle}{\pi R_{ev}^2} = \frac{2}{k^2 R_{ev}^2} \mathrm{Tr}(\hat{T}^{+}\hat{T}) = \frac{2}{k^2 R_{ev}^2} \sum_{i,j=1}^{2} \sum_{n,n'=1}^{\infty} \sum_{m=-n}^{n} \sum_{m'=-n'}^{n'} | T_{mn,m'n'}^{ij} |^2 . \quad \text{(S22)}$$

For particles with axial symmetry, averaging of cross sections over orientations can also be performed numerically using the integral (see Figure S5)

$$\langle Q_{ext} \rangle = \int_0^{\pi/1} \sin \vartheta d\vartheta \frac{1}{2} \left[ Q_{ext}^{TE}\left(\vartheta, A_s^{TE}(\vartheta)\right) + Q_{ext}^{TM}\left(\vartheta, A_s^{TM}(\vartheta)\right) \right], \quad \text{(S23)}$$

in which we have explicitly indicated that the normalized cross sections $Q_{ext}^{TE}\left(\vartheta\right)$ and $Q_{ext}^{TM}\left(\vartheta\right)$ themselves depend on the orientation angle $\vartheta$ even when the scattering coefficient $A_s$ is held constant. Furthermore, the cross sections depend on the scattering coefficients $A_s^{TE}(\vartheta)$ and $A_s^{TM}(\vartheta)$, which must be computed for each particle orientation. Consequently, in computing the average, it is necessary to solve the nonlinear pair of equations (17) and (18) from the main text for every orientation.

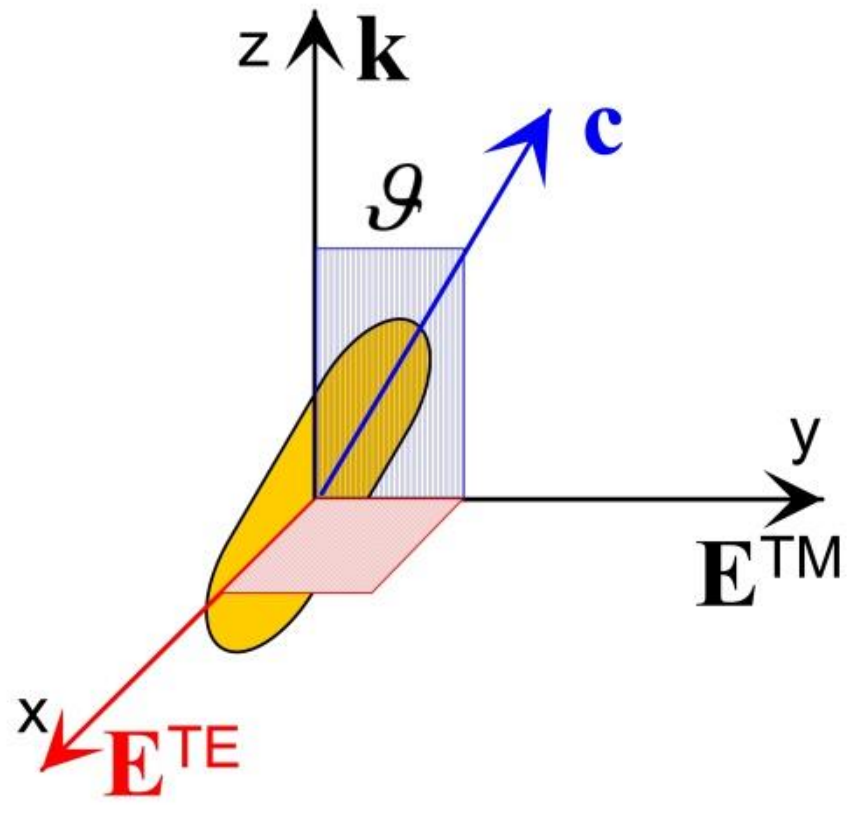

**Figure S5**. Schematic diagram for computing the orientation-averaged cross section. The vector $\mathbf{c}$ denotes the symmetry axis of the particle in the (y,z) plane at an angle $\vartheta$ with respect to the wave vector $\mathbf{k}$.

When employing numerical methods such as FEM, BEM (COMSOL), or DDSCAT, DDA, orientation averaging is typically performed by computing the mean cross section for a specified set of $N$ configurations of wave propagation directions $k_\alpha$ and polarizations $E_\beta$ [15, 18]

$$\langle C\rangle = \frac{1}{N}\sum_{\alpha,\beta} C_{\alpha,\beta} \,. \tag{S24}$$

Given that, for the TE mode, the cross section exhibits weak dependence on the orientation angle (see below), we may replace the exact averaging in (S17) with the approximate relation

$$\langle Q_{ext}\rangle = \frac{\frac{1}{2}\left[Q_{ext}^{TE}\left(0, A_s^{TE}(0)\right) + Q_{ext}^{TE}\left(90, A_s^{TE}(90)\right)\right]}{2} + \frac{1}{2}\int_0^{\pi/2} \sin\vartheta d\vartheta \left[Q_{ext}^{TM}\left(\vartheta, A_s^{TM}(\vartheta)\right)\right] \tag{S25}$$

where the corresponding scattering coefficient $A_s$, obtained from the iterative solution, is used for each mode. This raises an important question: to what extent is the substitution of the exact integral (S17) by the approximate expression (S19) justified? For dielectric particles [19] and gold nanorods [20], it has been demonstrated that such a replacement is admissible without appreciable loss of accuracy. Figure S6 presents the orientation-angle dependence of the extinction cross section for silver particles under TE polarization. A diameter of 10 nm and an axial ratio of $AR = 6$ correspond to the parameters of the particles investigated in the present work. The calculations were performed at wavelengths corresponding to extinction peaks 1–4 in the spectrum of randomly oriented particles (Figure S6, panel G).

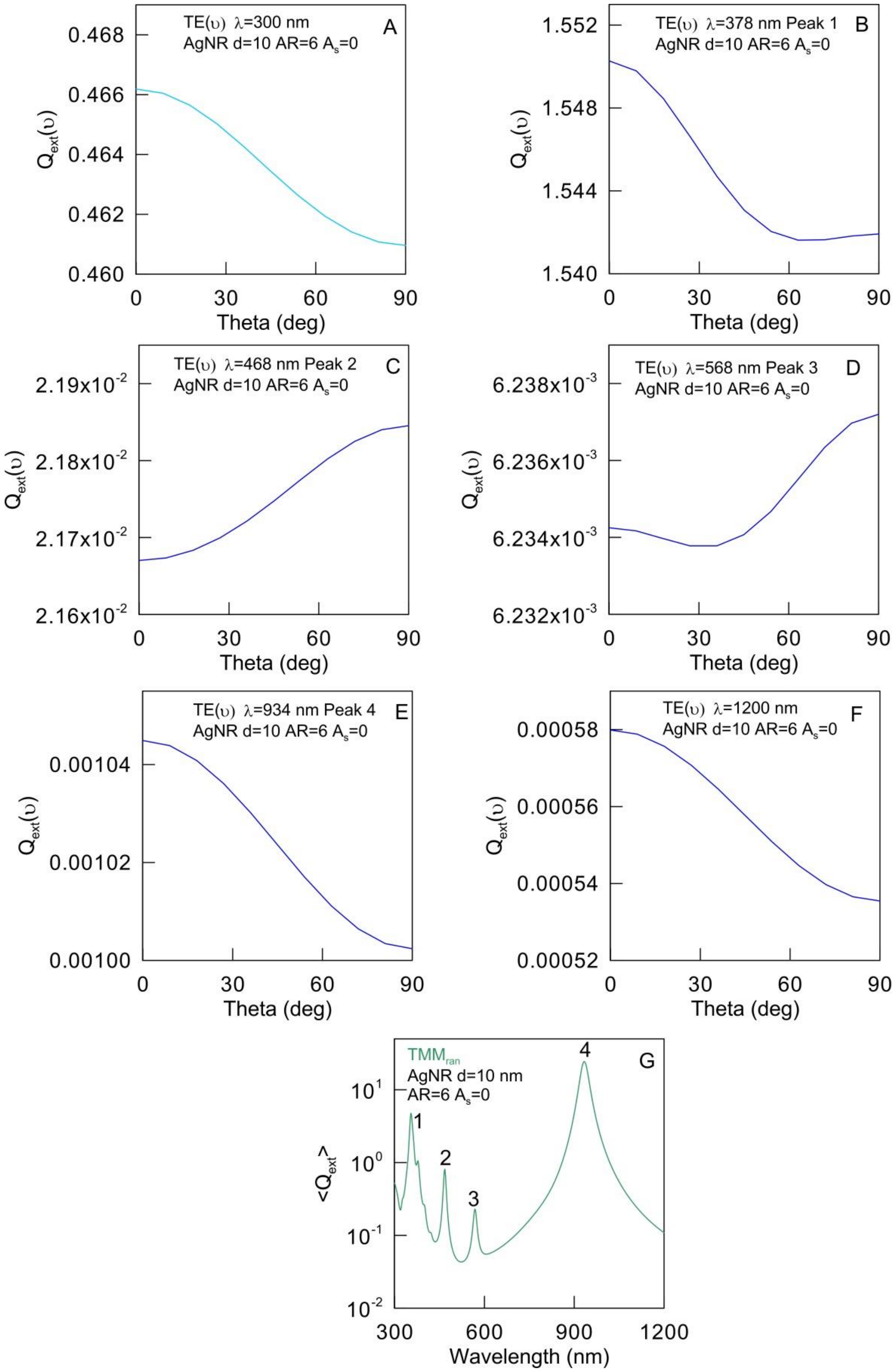

**Figure S6**. Dependence of the TE extinction cross section of a silver nanorod ( $d = 10$ nm, $AR = 6$ ) on the orientation angle $\vartheta$ (see Fig. S6) between the wave vector $\mathbf{k}$ and the symmetry axis $\mathbf{a}$ . Peak numbers 1–4 correspond to the successive peaks in the extinction spectrum of randomly oriented particles (panel G).

It is noteworthy that the angular dependence changes its character between peaks 1–2 and peaks 3–4. However, it is evident that the dependences are fairly monotonic, and the variations in values occur predominantly in the third to fourth significant digits. It is readily apparent that the half-sum will approximate the average to approximately an order of magnitude better. The most "pronounced" dependence is observed for 1200 nm, but there the values of the efficiency factor are so small that the error in the overall average cross section (S19) will be negligible.

Figures S7A and B compare the normalized numerical orientation-averaged cross section ($COM_{av}$, green line):

$$\langle Q_{ext} \rangle = \langle C_{ext} \rangle / \pi R_{ev}^2 = \frac{Q_{(k_x,E_y)} + Q_{(k_x,E_z)} + Q_{(k_y,E_x)} + Q_{(k_y,E_z)} + Q_{(k_z,E_x)} + Q_{(k_z,E_y)}}{6} = \\ = \frac{Q_{(k_x,E_y)} + Q_{(k_x,E_z)} + Q_{(k_z,E_y)}}{3} \cong \frac{2Q_{(k_x,E_y)} + Q_{(k_x,E_z)}}{3} \qquad \text{(S26)}$$

with the analytical T-matrix orientation-averaged cross section (Eq. (S21), blue line, TMM). In Eq. (S26), the rotation axis of the nanorods is assumed to be oriented along the z-axis, and some obvious symmetry relations have been applied. The last approximate relation is valid only for small particles, when $Q_{(k_x,E_y)} \cong Q_{(k_z,E_y)} = Q_{(k_z,E_x)}$. We note the absence of the quadrupole peak near 375 nm in the numerically averaged absorption spectrum obtained using Eq. (26) (green line). If orientation averaging is performed using the integral equation (22) of the main text, the quadrupole peak appears and the COMSOL (blue line) and TMM (green line) spectra agree almost perfectly (see panels C and D for a short-wavelength portion of the spectra). In fact, the COMSOL and TMM absorption and scattering spectra are indistinguishable except for a small peak near 375 nm (see the zoomed insets in panels C and D). Because of the complex multipolar features of the short-wavelength spectrum of silver rods, we had to use extended-precision T-matrix codes for aspect ratios greater than 4. Even with these codes, the convergence of the calculated cross-sections was slow because of the extremely high T-matrix order of about 40.

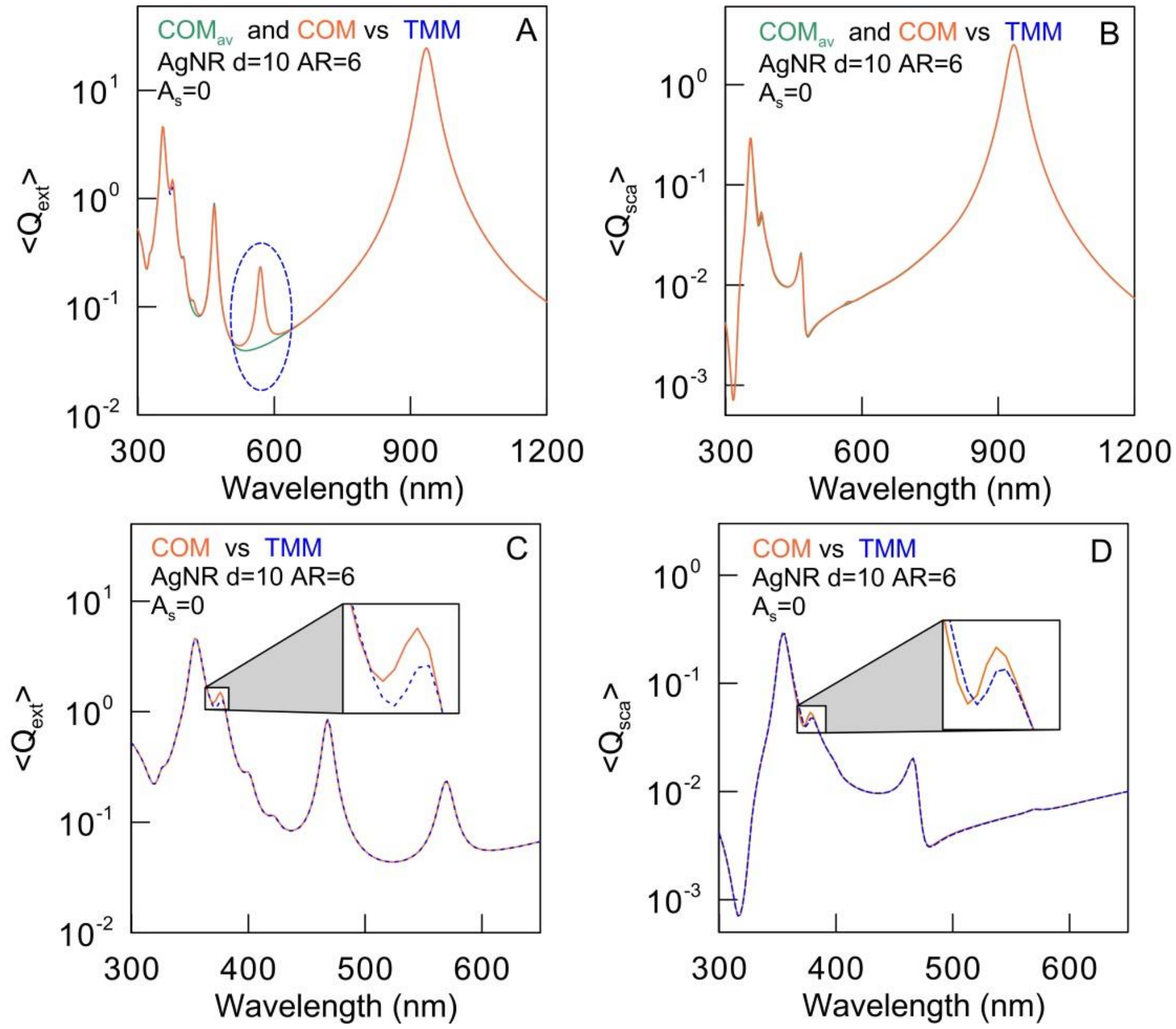

**Figure S7**. Extinction (A) and scattering (B) spectra calculated by approximate averaging of several scattering configurations (Eq. (S26), green lines), by the integral relation given by Eq. (22) of the main text (COM, orange lines), and by analytical T-matrix averaging (TMM, blue lines). The dashed blue ellipse marks the quadrupole peak (it is not seen in the green Eq. (S26) spectrum). Panels C and D show a short-wavelength portion of the absorption (C) and scattering (D) spectra. The zoomed insets show a small difference between the COM and TMM spectra near the weak 375 nm peak.

## Section S6. Comparison of the surface scattering constants for nanorods and spheroids

Figure S5 compares the principal tensor components $A_s^z$ and $A_s^x$ calculated numerically for nanorods and analytically for spheroids with the same diameters and aspect ratios of 2, 4, and 6. Large differences are observed near the resonance peaks for the tensor component $A_s^z$ of the nanorods, whereas the plots for the spheroids show a monotonic increase. For perpendicular excitation (panel B), the plots for the nanorods and spheroids exhibit similar behavior. The inset in panel C illustrates a typical weak spectral dependence of the electron scattering constants at wavelengths above 700 nm.

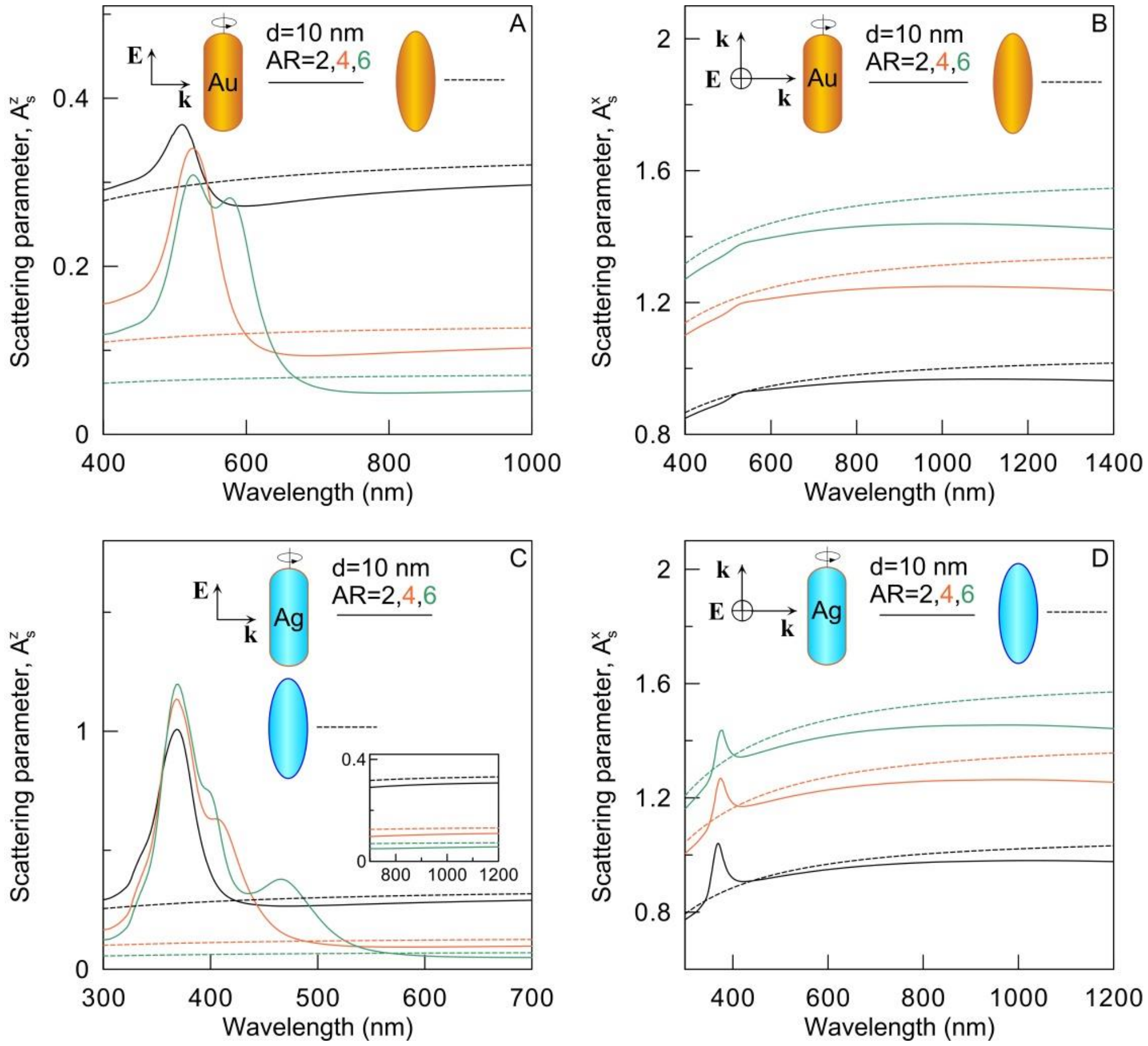

**Figure S8**. Spectral dependence of the principal components of the scattering tensor, $A_s^z$ (A, C) and $A_s^x$ (B, D), computed for gold (A, B) and silver (C, D) nanorods (solid lines) and compared with those for spheroids (dashed lines). Particle diameters and aspect ratios are given in the panels. The inset shows the long-wavelength region of the plots.

## Section S7. Additional spectra for gold and silver nanorods

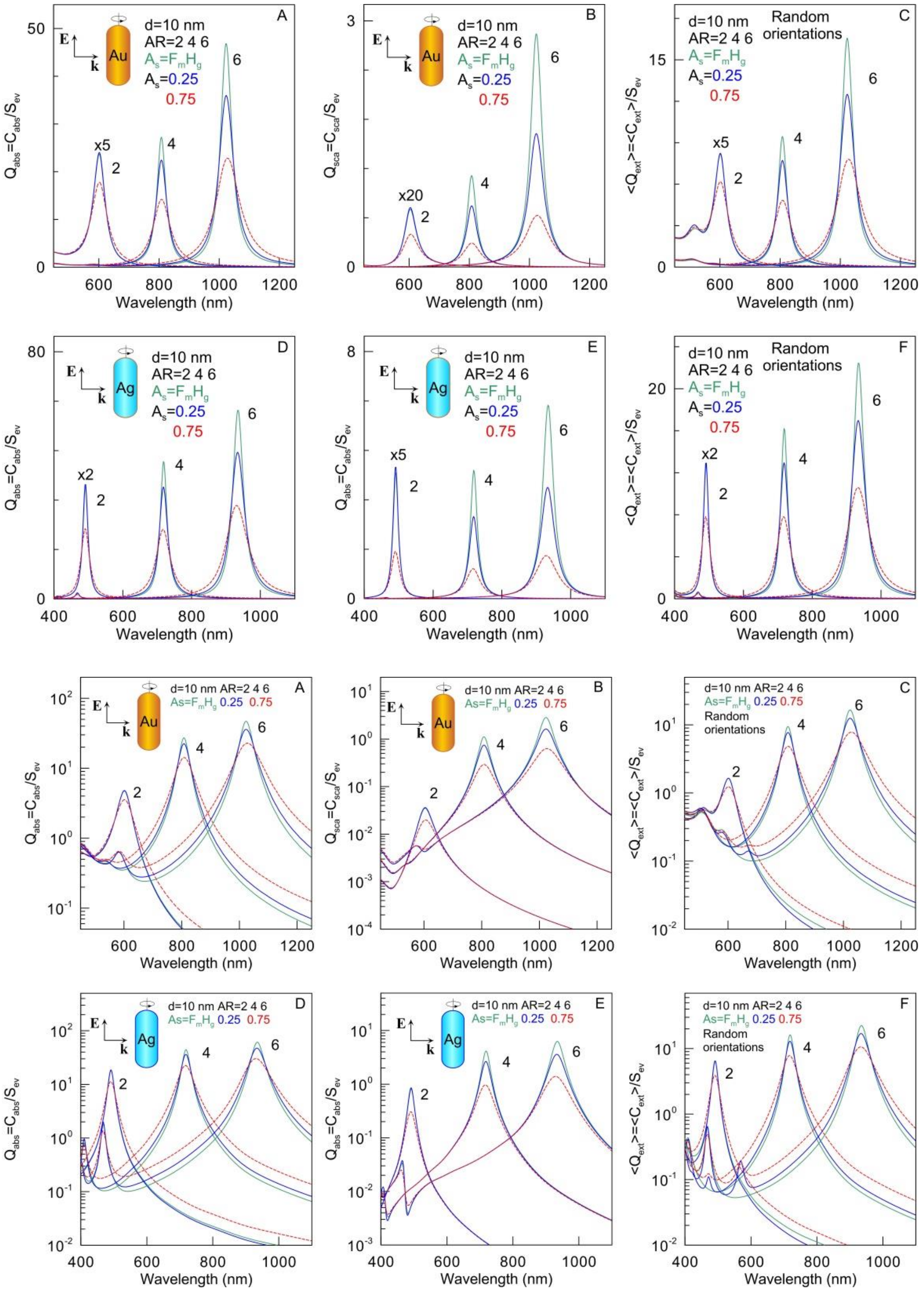

**Figure S9.** Absorption (A, D), scattering (B, E), and extinction (C, F) spectra of gold (A–C) and silver (D–F) nanorods calculated for longitudinal excitation (A, B, D, E) and random orientations (C, F). Green lines: COMSOL; blue and red lines: TMM. Nanorod parameters are given in the panels. Upper two rows: linear ordinate scale; bottom two rows: logarithmic ordinate scale.

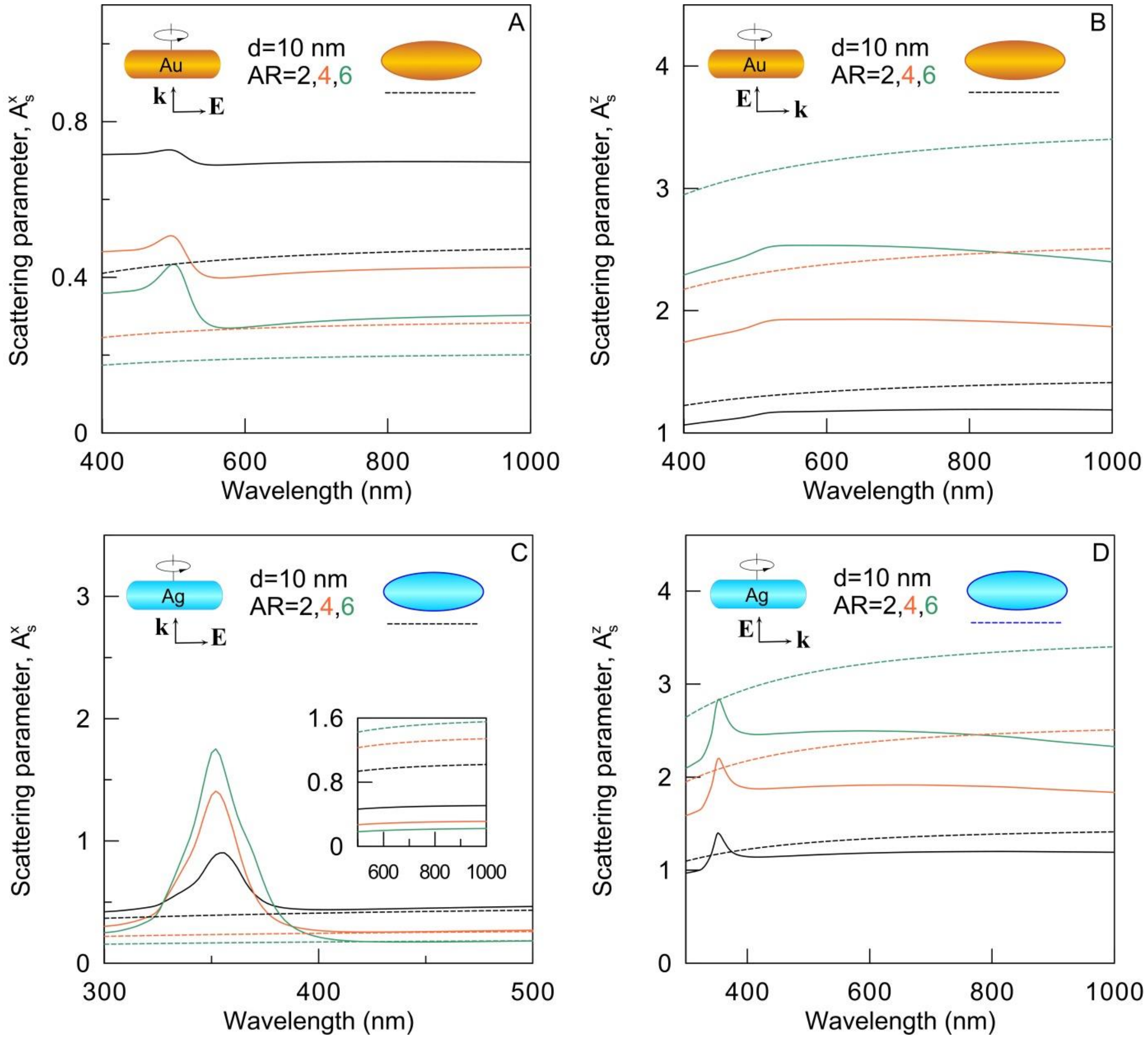

**Figure S10**. Spectral dependence of the principal components of the scattering tensor, $A_s^x$ (A, C) and $A_s^z$ (B, D), computed for gold (A, B) and silver (C, D) nanodisks (solid lines) and compared with those for spheroids (dashed lines). The disk thickness is 10 nm; the aspect ratios are $AR = 2$ (black lines), 4 (orange), and 6 (green).

## Section S8. Integral contributions to the surface-scattering broadening parameter from the flat and edge parts of the disk surface

The disk diameter, thickness, and aspect ratio are $D$, $t$, and $AR = D/t$, respectively (Figure S1). The disk volume and surface consist of the cylindrical (1) part

$$V_1 = \frac{\pi}{4}(D-t)^2 t = \frac{\pi}{4} t^3 \left(AR-1\right)^2, \; S_1 = 2\frac{\pi(D-t)^2}{4} = \frac{\pi}{2} t^2 \left(AR-1\right)^2 \tag{S27}$$

and the edge (2) part of the volume and surface [21]

$$V_2 = \frac{\pi}{6} t^3 \left[1 + \frac{3\pi}{4}\left(AR-1\right)\right], \; S_2 = \pi t^2 \left[1 + \frac{AR-1}{2}\pi\right]. \tag{S28}$$

For a disk with $t = 10$ nm and $AR = D/t = 4$ we have $V_1 = 7069\,\text{nm}^3$, $S_1 = 1414\,\text{nm}^2$ , $V_2 = 4225\,\text{nm}^3$, $S_2 = 1795\,\text{nm}^2$. Eq. (12) of the main text can be rewritten as

$$H_g^{\mathbf{e}}(\omega) = \frac{a_0 \int_S \left|\mathbf{n}\mathbf{E}_i\right|^2 \mathrm{d}S}{\int_V \left|\mathbf{E}_i\right|^2 \mathrm{dV}} = \frac{a_0 S F_s}{V F_V}, \tag{S29}$$

with

$$F_s = \frac{1}{S}\int_S \left|\mathbf{n}\mathbf{E}_i\right|^2 \mathrm{d}S, \tag{S30}$$

$$F_V = \frac{1}{V}\int_V \left|\mathbf{E}_i\right|^2 \mathrm{dV}. \tag{S31}$$

The surface and volume integrals (S30) and (S31) can be represented as the sums of cylindrical (1) and edge (2) parts of the disks

$$F_s = \frac{S_1}{S}\frac{1}{S_1}\int_{S_1} \left|\mathbf{n}\mathbf{E}_i\right|^2 \mathrm{d}S + \frac{S_2}{S}\frac{1}{S_2}\int_{S_2} \left|\mathbf{n}\mathbf{E}_i\right|^2 \mathrm{d}S = \frac{S_1}{S}F_{S_1} + \frac{S_2}{S}F_{S_2} \equiv F_s^1 + F_s^2, \tag{S32}$$

$$F_V = \frac{V_1}{V}\frac{1}{V_1}\int_{V_1} \left|\mathbf{E}_i\right|^2 \mathrm{dV} + \frac{V_2}{V}\frac{1}{V_2}\int_{V_2} \left|\mathbf{E}_i\right|^2 \mathrm{dV} = \frac{V_1}{V}F_{V_1} + \frac{V_2}{V}F_{V_2} \equiv F_V^1 + F_V^2. \tag{S32}$$

The two contributions to the surface and volume integrals, $F_S^i$ and $F_V^i$ ($i = 1, 2$), are determined by the specific distribution of the internal field and its normal component over the partial volumes and surfaces 1 and 2 (through the functions $F_{S_i}$ and $F_{V_i}$) and by the surface and volume fractions $S_i / S$ and $V_i / V$, $i = 1, 2$). Notably, the flat-surface contribution $F_s^1$ in panel A is comparable to the corresponding edge contribution, even though the exciting field is parallel to the flat surface. This clearly indicates multipole contributions generated by the rounded disk edge. The plot in panel C shows the 354-nm resonance of the ratio $F_S / F_V$, which coincides with the geometrical factor of $H_g^{\mathbf{e}}$ up to the normalization factor $a_0 S / V$ ($a_0 = \sqrt[3]{3V / 4\pi}$).

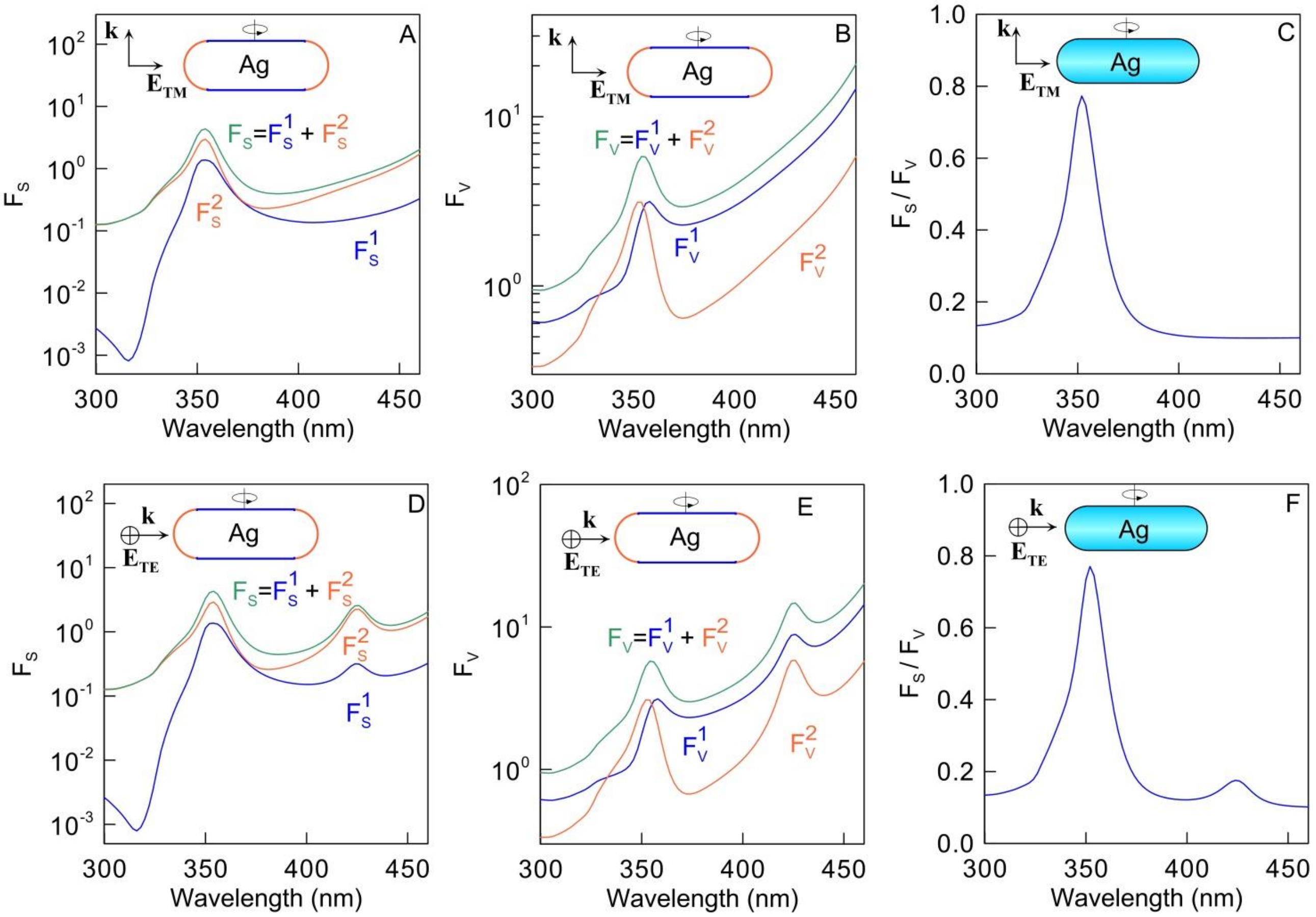

**Figure S11**. UV spectra of the integral contributions to the surface-averaged normal component of the inner field at the particle boundary (A, D) and to the volume-averaged inner field (B, E) for the cylindrical flat (superscript 1, blue lines) and rounded-edge (superscript 2, orange lines) parts under $TE_0 = TM_0$ (A-C) and $TE_{90}$ (D-F) excitations. Green lines represent the total summed quantities. Panels C and D show the UV spectrum of the total $F_S / F_V$ ratio, which is proportional to the $H_g^{\mathbf{e}}$ geometrical factor (see Eq. (12) in the main text). Calculations were performed for a silver disk with a thickness of 10 nm and aspect ratio of 4. Note the appearance of the quadrupole peaks at 425 nm under TE90 excitation when both **k** and **E** vectors have components in the disk plane

## Section S9. Material functions $1/|\varepsilon_r + 1|$ for gold and silver in water

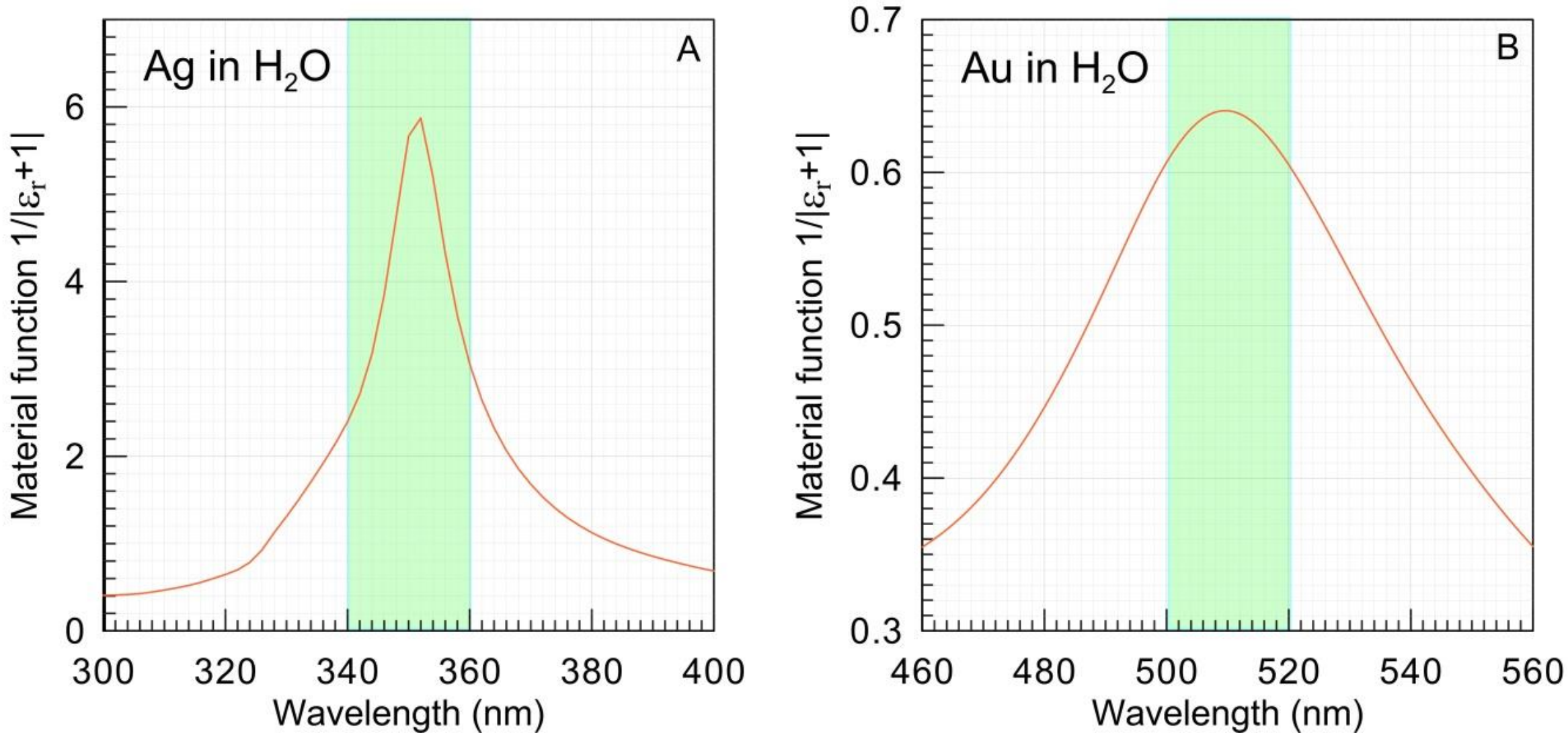

Figure S12. UV spectra of the material function $1/|\varepsilon_r + 1|$ for gold and silver in water.

## Section S10. The optical dielectric function of bulk silver

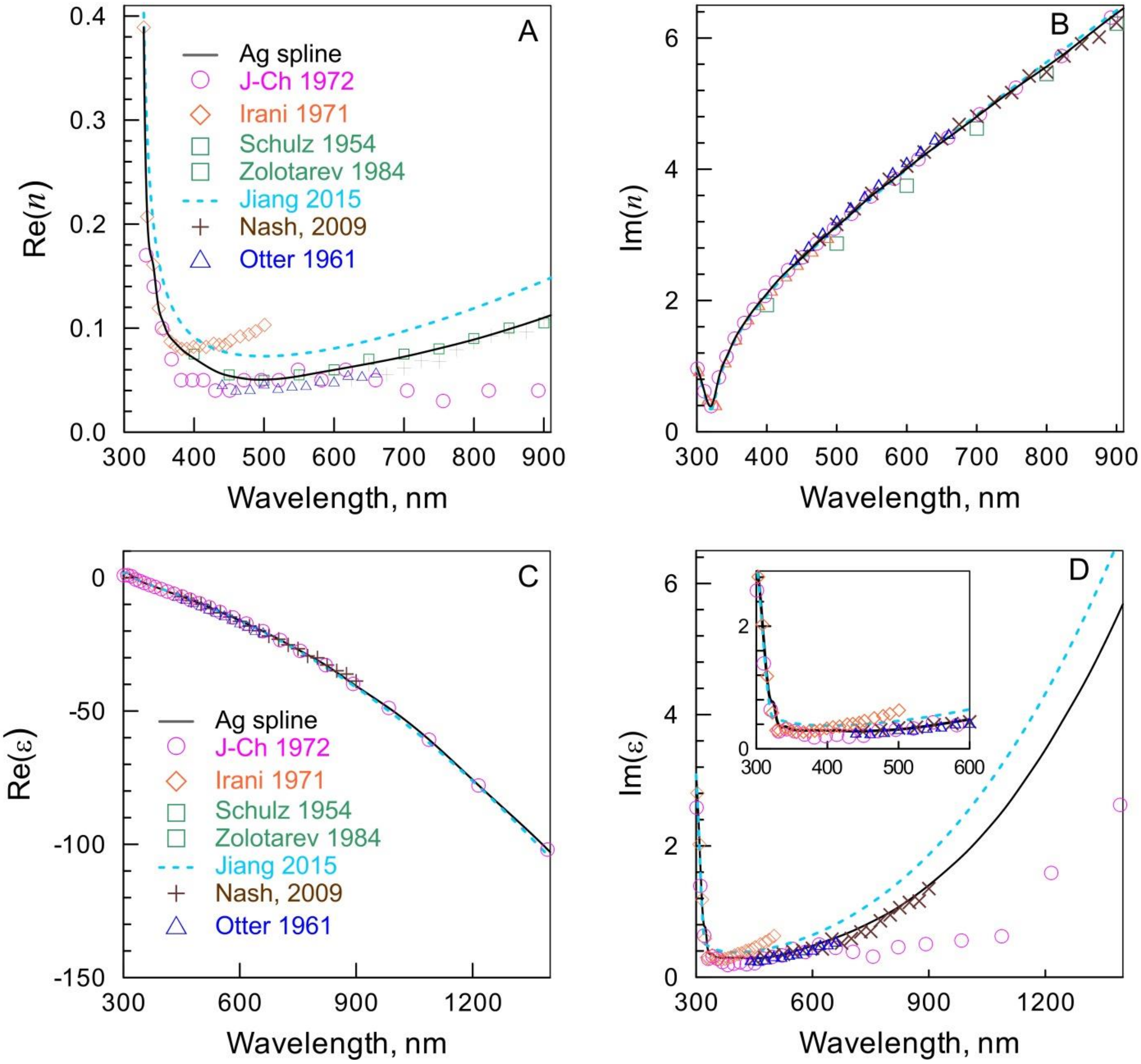

**Figure S13**. Spectral dependence of the bulk Ag refractive index $n$ (A, B) and dielectric function $\varepsilon$ (C, D) taken from Refs. [8-26] or recalculated by $\varepsilon = n^2$. The black solid line shows the cubic spline approximation (see below).

The dielectric function of water was calculated by the equation

$$\varepsilon_m = \left[1.3233 + \frac{3.478\times10^{-3}}{\lambda^2} - \frac{5.111\times10^{-5}}{\lambda^4}\right]^2, \ [\lambda] = \mu\text{m}\,. \quad \text{(S27)}$$

**FORTRAN subroutine for spline-approximation of the bulk Ag optical constants:**

```
      subroutine Ag2021(wavel,ref,eps)
      IMPLICIT REAL*8(A-H,O-Z)
c     wavel is the wavelength (in nm), ref and eps are the complex bulk constants
      REAL*8 MN,MK
      DIMENSION XN(100),YN(100),MN(100),XK(100),YK(100),MK(100)
      complex*16 ref,eps,im
c
      xx=wavel
      if(xx.gt.600) go to 1

c     Spline G1-4 parameters
       XN(1)=  299.5
       XN(2)=  302.9
       XN(3)=  305.2
       XN(4)=  308.6
       XN(5)=  312.8
       XN(6)=  315.3
       XN(7)=  318.7
       XN(8)=  322.7
       XN(9)=  325.7
       XN(10)=  328
       XN(11)=  330
       XN(12)=  333
       XN(13)=  336.8
       XN(14)=  341.5
       XN(15)=  345.9
       XN(16)=  349.4
       XN(17)=  355
       XN(18)=  360
       XN(19)=  370
       XN(20)=  380
       XN(21)=  400
       XN(22)=  420
       XN(23)=  440
       XN(24)=  460
       XN(25)=  480
       XN(26)=  500
       XN(27)=  520
       XN(28)=  540
       XN(29)=  560
       XN(30)=  580
       XN(31)=  600
       XN(32)=  625
       XN(33)=  650
       XN(34)=  675
       XN(35)=  700
       XN(36)=  750
       XN(37)=  800
       XN(38)=  850
       XN(39)=  900
       XN(40)=  950
       XN(41)=  1000
       XN(42)=  1050
       XN(43)=  1100
       XN(44)=  1150
       XN(45)=  1200
       XN(46)=  1250
       XN(47)=  1300
       XN(48)=  1350
       XN(49)=  1400
       XN(50)=  1450
       XN(51)=  1500
       XN(52)=  1550
       XN(53)=  1600
C
       YN(1)=  1.564
       YN(2)=  1.556
       YN(3)=  1.543
       YN(4)=  1.47
       YN(5)=  1.343
       YN(6)=  1.248
       YN(7)=  1.083
```

YN(8)=  0.877
YN(9)=  0.622
YN(10)=  0.389
YN(11)=  0.282
YN(12)=  0.207
YN(13)=  0.174
YN(14)=  0.161
YN(15)=  0.136
YN(16)=  0.119
YN(17)=  0.107
YN(18)=  0.098
YN(19)=  0.088
YN(20)=  0.081
YN(21)=  0.071
YN(22)=  0.0625
YN(23)=  0.056
YN(24)=  0.0528
YN(25)=  0.0508
YN(26)=  0.0505
YN(27)=  0.0512
YN(28)=  0.0528
YN(29)=  0.055
YN(30)=  0.0575
YN(31)=  0.0598
YN(32)=  0.0628
YN(33)=  0.0659
YN(34)=  0.0691
YN(35)=  0.0725
YN(36)=  0.0804
YN(37)=  0.089
YN(38)=  0.099
YN(39)=  0.11
YN(40)=  0.1225
YN(41)=  0.1355
YN(42)=  0.15
YN(43)=  0.165
YN(44)=  0.182
YN(45)=  0.199
YN(46)=  0.218
YN(47)=  0.237
YN(48)=  0.258
YN(49)=  0.28
YN(50)=  0.303
YN(51)=  0.3265
YN(52)=  0.352
YN(53)=  0.378
C
MN(1)=  -9.877E-005
MN(2)=  -9.877E-005
MN(3)=  -0.007971
MN(4)=  -0.001121
MN(5)=  -0.002014
MN(6)=  -0.00595
MN(7)=  0.00355
MN(8)=  -0.01253
MN(9)=  -0.01325
MN(10)=  0.03488
MN(11)=  0.008653
MN(12)=  0.0049
MN(13)=  0.001394
MN(14)=  -0.001449
MN(15)=  0.0005294
MN(16)=  0.0008458
MN(17)=  -0.0001715
MN(18)=  0.0001913
MN(19)=  -8.268E-006
MN(20)=  2.173E-005
MN(21)=  -1.057E-006
MN(22)=  4.999E-006
MN(23)=  1.106E-005
MN(24)=  2.58E-007
MN(25)=  5.907E-006
MN(26)=  1.613E-006
MN(27)=  2.642E-006
MN(28)=  1.318E-006
MN(29)=  1.087E-006
MN(30)=  -1.166E-006
MN(31)=  5.777E-007

```
            MN(32)=  5.315E-008
            MN(33)=  1.697E-007
            MN(34)=  2.281E-007
            MN(35)=  8.379E-007
            MN(36)=  1.233E-008
            MN(37)=  7.928E-007
            MN(38)=  1.765E-007
            MN(39)=  9.014E-007
            MN(40)=  -1.82E-007
            MN(41)=  1.027E-006
            MN(42)=  -3.245E-007
            MN(43)=  1.471E-006
            MN(44)=  -7.612E-007
            MN(45)=  1.573E-006
            MN(46)=  -7.324E-007
            MN(47)=  1.356E-006
            MN(48)=  1.074E-007
            MN(49)=  6.143E-007
            MN(50)=  -1.644E-007
            MN(51)=  1.243E-006
            MN(52)=  -8.651E-009
            MN(53)=  -8.651E-009
1           continue
C
           XK(1)=  248.9
           XK(2)=  255.2
           XK(3)=  261.6
           XK(4)=  268.9
           XK(5)=  276.1
           XK(6)=  284.4
           XK(7)=  292.5
           XK(8)=  300.9
           XK(9)=  310.7
           XK(10)=  320.4
           XK(11)=  331.5
           XK(12)=  342.5
           XK(13)=  354.3
           XK(14)=  367.9
           XK(15)=  381.5
           XK(16)=  397.4
           XK(17)=  413.3
           XK(18)=  430.5
           XK(19)=  450.9
           XK(20)=  471.4
           XK(21)=  496
           XK(22)=  521
           XK(23)=  548.6
           XK(24)=  582.1
           XK(25)=  616.9
           XK(26)=  659.5
           XK(27)=  704.5
           XK(28)=  756.1
           XK(29)=  821.2
           XK(30)=  892
            XK(31)=  984.1
            XK(32)=  1087.7
            XK(33)=  1215.5
            XK(34)=  1393.2
           XK(35)=1610.2
C
           YK(1)= 1.38900E+00
           YK(2)= 1.39300E+00
           YK(3)= 1.38700E+00
           YK(4)= 1.37200E+00
           YK(5)= 1.33100E+00
           YK(6)= 1.26400E+00
           YK(7)= 1.16100E+00
           YK(8)= 9.64000E-01
           YK(9)= 6.16000E-01
           YK(10)=3.92000E-01
           YK(11)=8.29000E-01
           YK(12)=1.14200E+00
           YK(13)=1.41900E+00
           YK(14)=1.65700E+00
           YK(15)=1.86400E+00
           YK(16)=2.07000E+00
           YK(17)=2.27500E+00
           YK(18)=2.46200E+00
```

```
      YK(19)=2.65700E+00
      YK(20)=2.86900E+00
      YK(21)=3.09300E+00
      YK(22)=3.32400E+00
      YK(23)=3.58600E+00
      YK(24)=3.85800E+00
      YK(25)=4.15200E+00
      YK(26)=4.48300E+00
      YK(27)=4.83800E+00
      YK(28)=5.24200E+00
      YK(29)=5.72700E+00
      YK(30)=6.31200E+00
      YK(31)=6.99200E+00
      YK(32)=7.79500E+00
      YK(33)=8.82800E+00
      YK(34)=1.01000E+01
      YK(35)=1.18500E+01
C
       MK(1)=  -0.0003008081
       MK(2)=  -0.0003008081
       MK(3)=  1.580026E-005
       MK(4)=  -0.0007139049
       MK(5)=  -0.0001736024
       MK(6)=  -0.0004512451
       MK(7)=  -0.001434666
       MK(8)=  -0.001597511
       MK(9)=  -0.0002189577
       MK(10)=  0.01017516
       MK(11)=  -0.004179221
       MK(12)=  0.0005716763
       MK(13)=  -0.0008454766
       MK(14)=  2.622849E-005
       MK(15)=  -0.0002650559
       MK(16)=  0.0001065387
       MK(17)=  -0.0001848276
       MK(18)=  -9.212307E-005
       MK(19)=  0.0001091794
       MK(20)=  -0.0001149184
       MK(21)=  2.897708E-005
       MK(22)=  3.033301E-005
       MK(23)=  -8.691511E-005
       MK(24)=  4.607977E-005
       MK(25)=  -4.050878E-005
       MK(26)=  1.402204E-005
       MK(27)=  -3.861147E-007
       MK(28)=  -1.769247E-005
       MK(29)=  2.877102E-005
       MK(30)=  -2.531625E-005
       MK(31)=  1.014666E-005
       MK(32)=  5.466382E-006
       MK(33)=  -1.243508E-005
       MK(34)=  7.599362E-006
       MK(35)=  7.599362E-006
c--------------------------------------------------
      if(xx.gt.600.) then
        refre=0.0755 - 0.000155*xx + 2.15E-7*xx*xx
        go to 29
      end if
C     Step1 The refractive index calculations
      J=2
10    IF (XX.LE.XN(J)) GO TO 20
      J=J+1
      GO TO 10
20    I=J
      XN0=XN(I)-XX
      XN1=XX-XN(I-1)
      H=XN(I)-XN(I-1)
      H2=H*H/6.0
      refre=MN(I-1)*XN0*XN0*XN0/6./H+MN(I)*XN1*XN1*XN1/6./H+
     c  (YN(I-1)-MN(I-1)*H2)*XN0/H+(YN(I)-MN(I)*H2)*XN1/H
29    J=2
30    IF (XX.LE.XK(J)) GO TO 40
      J=J+1
      GO TO 30
40    I=J
      XN0=XK(I)-XX
      XN1=XX-XK(I-1)
      H=XK(I)-XK(I-1)
```

```
      H2=H*H/6.0
      refim=MK(I-1)*XN0*XN0*XN0/6./H+MK(I)*XN1*XN1*XN1/6./H+
    c (YK(I-1)-MK(I-1)*H2)*XN0/H+(YK(I)-MK(I)*H2)*XN1/H
C     ----------------------------------------------------------------
C     Step2: calculation of the bulk parameter eps=eps1+ieps2
      eps1=refre*refre-refim*refim
      eps2=2.d0*refre*refim
      im=(0.d0,1.d0)
      eps=eps1+im*eps2
      ref=refre+im*refim
c     --------------------------------------------------------------------
      RETURN
      END
```